\documentclass[12pt,showlabels]{article}
 
\usepackage{amsmath,amssymb,amsthm}
\usepackage{graphicx}
\usepackage{ifpdf}
\usepackage{color}

\allowdisplaybreaks
\DeclareGraphicsExtensions{.pdf,.png,.jpg,.jpeg,.mps,.eps}

\usepackage[all]{xypic}
\usepackage[colorlinks=false,linktocpage=true]{hyperref}

\usepackage[all]{xypic}
\usepackage{hyperref}
\hypersetup{colorlinks=false}
    
\newtheorem{theorem}{Theorem}[section]

\newtheorem{definition}[theorem]{Definition}

\newtheorem{remark}[theorem]{Remark}

\newtheorem{corollary}[theorem]{Corollary}
\newtheorem{lemma}[theorem]{Lemma}

\def\R{\mathbb R}

\def\N{\mathbb N}
\def\Z{\mathbb Z}
\def\G{\mathsf{G}}
\def\H{\mathbb H}

\def\X{\mathcal X}

\def\P{{\mathcal P}}

\newcommand{\eps}{\varepsilon}
\def\tr{{\rm tr}}
\newcommand{\PSL}{{\rm PSL}}
\newcommand{\SL}{{\rm SL}}
\def\arccosh{{\rm arccosh}}

\begin{document}

\title{Partner orbits and action differences on compact factors of the hyperbolic plane. 
Part I: Sieber-Richter pairs}
\author{{\sc H.M.~Huynh$^{1,\,2}$ \& M.~Kunze$^{2}$} \\[1ex]
       $^{1}$ Department of Mathematics, Quy Nhon University \\
       Binh Dinh, Vietnam \\[1ex]
       $^{2}$ Universit\"at K\"oln, Institut f\"ur Mathematik \\
       Weyertal 86-90, D\,-\,50931 K\"oln, Germany \\[1ex]
       e-mail: hhien@mi.uni-koeln.de\\
       \quad\hskip3.4em mkunze@mi.uni-koeln.de
       }

\maketitle        

\begin{abstract}
\noindent
Physicists have argued that periodic orbit bunching
leads to universal spectral fluctuations for chaotic quantum systems.
To establish a more detailed mathematical understanding of this fact, 
it is first necessary to look more closely at the classical side of the problem 
and determine orbit pairs consisting of orbits which have similar actions. 
In this paper we specialize to the geodesic flow on compact factors of the hyperbolic plane 
as a classical chaotic system. 
We prove the existence of a periodic partner orbit for a given periodic orbit 
which has a small-angle self-crossing in configuration space which is a `2-encounter'; 
such configurations are called `Sieber-Richter pairs' in the physics literature. 
Furthermore, we derive an estimate for the action difference of the partners. 
In the second part of this paper \cite{partII}, an inductive argument is provided 
to deal with higher-order encounters. 
\end{abstract}


\setcounter{equation}{0}

\section{Introduction}

In the semi-classical limit chaotic quantum systems very often exhibit universal behavior,
in the sense that several of their characteristic quantities agree with the respective
quantities found for certain ensembles of random matrices. Via trace formulae, such
quantities can be expressed as suitable sums over the periodic orbits of the underlying
classical dynamical system. For instance, the two-point correlator function is 
\begin{equation}\label{formfactor}
   K(\tau)=\Big\langle \frac{1}{T_H}\sum_{\gamma,\gamma'}A_\gamma 
   A_{\gamma'}^* e^{\frac{i}{\hbar}(S_\gamma-S_{\gamma'})}
   \delta\Big(\tau T_H-\frac{T_\gamma+T_{\gamma'}}{2}\Big) \Big\rangle,
\end{equation}
where $\langle\cdot\rangle$ abbreviates the  average over the energy and
over a small time window, $T_H$ denotes the Heisenberg time and $A_\gamma$, $S_\gamma$, and $T_\gamma$ 
are the amplitude, the action, and the period of the orbit $\gamma$, respectively. 

The contribution of the terms $\gamma=\gamma'$ to (\ref{formfactor}) is called 
the `diagonal approximation' and was studied by Hannay/Ozorio de Almeida \cite{hannozdeal} 
and Berry \cite{berry} in the 1980's; a mathematical rigorous treatment is still missing. 
Also see \cite{KeatRob} for other work on the diagonal approximation. 

\begin{figure}[ht]
\begin{center}
\begin{minipage}[ht]{0.8\linewidth}
   \centering
   \includegraphics[angle=0,width=0.7\linewidth]{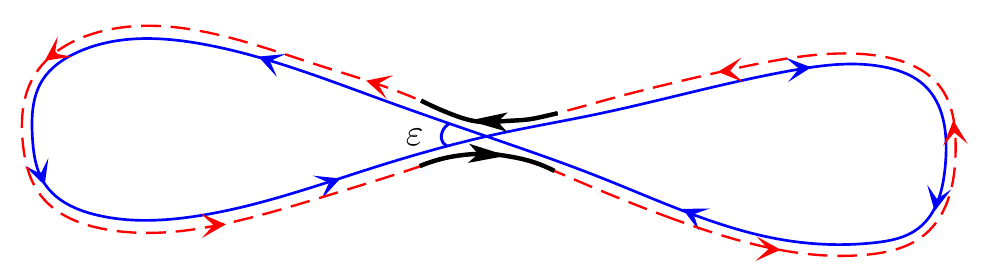}
\end{minipage}
\end{center}
\caption{Example of a Sieber-Richter pair}\label{existence}
\end{figure}

To next order, as $\hbar\to 0$, the main term  
from (\ref{formfactor}) arises owing to those orbit pairs $\gamma\neq\gamma'$ 
for which the action difference $S_\gamma-S_{\gamma'}$ is `small'. 
This was first considered by Sieber and Richter \cite{SieberRichter,Sieber2}, 
who argued that a given periodic orbit with a self-crossing in configuration space 
at a small angle $\eps$ (see Figure \ref{existence}) will admit a neighboring periodic orbit with almost the same action. 
Furthermore, $|S_\gamma-S_{\gamma'}|  \propto \eps^2$ was obtained for the action difference. 
The neighboring orbit is called a \textit{partner} of the given orbit 
and one calls the two orbits a \textit{Sieber-Richter pair}.

\noindent
In phase space, a Sieber-Richter pair contains a region where two stretches of the same orbit are almost mutually time-reversed
and one addresses this region as a {\em $2$-encounter} or, more strictly, a {\em $2$-antiparallel encounter};
the `2' stands for two orbit stretches which are close in configuration space, 
and `antiparallel' means that the two stretches have opposite directions; see Figure \ref{existence}. 
The two orbits noticeably differ from each other only by their connections inside the encounter region. 
The smaller is the crossing angle, the closer will be the stretches, and the longer the stretches remain close, 
the smaller will be the resulting action difference. Outside the encounter region, the two orbits are almost indistinguishable.  
In contrast to that, they practically coincide in one loop and appear as time-reversed in the other loop. 
This explains why Sieber-Richter pairs may only exist in systems which are invariant under time-reversal. 
Taking into account those contributions to (\ref{formfactor}), Sieber and Richter were able 
to calculate the first two terms in the expansion of (\ref{formfactor}) for small $\tau$ 
(the first term comes from the diagonal approximation), and it turned out that the result agreed 
with what is obtained using random matrix theory \cite{efetov}, for certain symmetry classes. 

This discovery prompted an increased research activity on the subject matter 
in the following years and finally led to an expansion 
\[ K(\tau)=2\tau-\tau\,\ln(1+2\tau)=2\tau-2\tau^2+2\tau^3+\ldots \] 
for the orthogonal ensemble (the symmetry class relevant for time-reversal invariant systems) 
to all orders in $\tau$, by including the higher-order encounters also; 
see \cite{mueller2004,mueller2005,mueller2009}, and in addition \cite{muellerthesis,haake}, 
which provide much more background and many further references.  
\medskip

To establish a more detailed mathematical understanding, it is natural to start, more modestly, 
on the classical side and try to prove the existence of partner orbits and derive good estimates for the action differences 
of the partners. For $2$-encounters this is done in the present work, where we consider 
the geodesic flow on factors of the hyperbolic plane; in this case the action of a periodic orbit is proportional to its length. 
The companion paper \cite{partII} then deals with the technically more involved higher-order encounters. 

In the physics community this system is often called the Hadamard-Gutzwiller model, 
and it has frequently been studied \cite{braun2002,Sieber1}; 
further related work includes \cite{haake,mueller2005,Turek05}.  
In Braun et al.~\cite{braun2002} symbolic dynamics was used to obtain an approximation for the length difference
$\Delta L \approx 4\ln(\cos(\eps/2))$, where $\eps$ denotes the crossing-angle; see also \cite{Sieber1}. 
In both papers there was no estimate on the error term given. For certain systems, 
including some with two degrees of freedom, Turek/Richter \cite{tu-ri} and Spehner \cite{Spehner} 
provided another approximation for the action difference of a Sieber-Richter pair 
by introducing suitable coordinates on the Poincar\'e section; 
also see M\"uller et al.~\cite{mueller2005} for related work. However, these coordinates were not calculated 
and once again there was no bound given for the error term of the action differences. 
This approach was also used for systems with higher degrees of freedom by Turek et al.~\cite{Turek05}. 

The paper is organized as follows. In section \ref{prel_sect} we introduce the necessary background 
material which is well-known in principle \cite{bedkeanser,einsward,KatHas}. Since the field requires a substantial amount 
of notation and identifications, we nevertheless provide some details. Then in section \ref{srp_sect} 
we turn to Sieber-Richter pairs. First we give a quantitative version of the shadowing lemma 
and the Anosov closing lemma; the latter may be of independent interest. Next we consider 
orbit self-crossings (in configuration space). Theorem \ref{existthm} contains 
a first main result about the existence of a partner orbit and an estimate for the action difference. 
If the crossing angle is small enough and the space is compact, then the partner is even unique; 
see theorem \ref{uniq_thm}. 
Finally we include a brief discussion of pseudo-orbits, i.e., of periodic orbits which 
decompose into smaller pieces. 
\medskip 

\noindent 
{\bf Acknowledgments:} This work was initiated in the framework 
of the collaborative research program SFB-TR 12 `Symmetries and Universality in Mesoscopic Systems' 
funded by the DFG, whose financial support is gratefully acknowledged. 
We enjoyed many fruitful discussions with P.~Braun, K.~Bieder, F.~Haake, G.~Knieper, and S.~M\"uller. 
Most results of this paper also appeared in the PhD thesis \cite{HMH} of the first author.  
The second author is grateful to C.~Series for sending him a copy of the book \cite{bedkeanser}.

\setcounter{equation}{0}

\section{Preliminaries}
\label{prel_sect} 

We consider the geodesic flow on compact Riemann surfaces of constant negative curvature. 
In fact this flow has had a great historical relevance for the development of the whole theory 
of hyperbolic dynamical systems or Anosov systems. It is well-known that any compact orientable surface 
with a metric of constant negative curvature is isometric to a factor $\Gamma\backslash \H^2$, 
where $\Gamma $ is a discrete subgroup of the projective Lie group $\PSL(2,\R)=\SL(2,\R)/\{\pm E_2\}$; 
here $\SL(2,\R)$ is the group of all real $2\times 2$ matrices with unity determinant, 
and $E_2$ denotes the unit matrix.  
General references for this section are \cite{bedkeanser,einsward,KatHas}, 
and these works may be consulted for the proofs to all results which are stated here without a proof. 

\subsection{$\H^2$ and $\PSL(2,\R)$} 

The hyperbolic plane is the upper complex half plane 
$\H^2=\{(x, y)\in\R^2: y>0\}$, endowed with the Riemannian metric $g={(g_z)}_{z\in\H^2}$ given by 
\[ g_z(\xi, \zeta)=\frac{1}{y^2}(\xi_1\zeta_1+\xi_2\zeta_2) \] 
for $z=(x, y)\in\H^2$, $\xi=(\xi_1, \xi_2)\in T_z(\H^2)$,
and $\zeta=(\zeta_1, \zeta_2)\in T_z(\H^2)$. 
The symbol $T_z(\H^2)$ denotes the tangent space to $\H^2$ at $z\in\H^2$ 
(which is $\R^2$). In short, $g=\frac{dx^2+dy^2}{y^2}$. 
Then $(\H^2, g)$ has constant curvature $-1$. 
The geodesics of $(\H^2, g)$ are given by infinite vertical lines and by semi-circles centered on the real axis. 

The M\"obius transformations ${\rm M\ddot ob}(\H^2)=\{z\mapsto \frac{az+b}{cz+d}\,:\, ad-bc =1\}$ 
can be identified with the projective group $\PSL(2,\R)=\SL(2,\R)/\{\pm E_2\}$ by means of the isomorphism
\begin{equation*}
   \Phi\Bigg(\pm\bigg(\begin{array}{cc}a&b\\c&d\end{array}\bigg)\Bigg)=z\mapsto\frac{az+b}{cz+d}.
\end{equation*} 
The unit tangent bundle of $\H^2$ is
\[ T^1\H^2=\{(z, \xi): z\in\H^2, \xi\in T_z\H^2,
   {\|\xi\|}_z=g_z(\xi, \xi)^{1/2}=1\}. \] 

For $g\in {\rm PSL}(2, \R)$ we define the derivative operator
${\cal D}g: T^1\H^2\to T^1\H^2$ by 
${\cal D}g(z, \xi)=(T(z), T'(z)\xi)$, 
where $T=\Phi(g)$. Explicitly, if $g=[G]=\{-G, G\}\in {\rm PSL}(2, \R)$ is the class generated 
by the matrix $G=\Big(\begin{array}{cc} a & b \\ c & d\end{array}\Big)\in {\rm SL}(2, \R)$, 
then $T(z)=\frac{az+b}{cz+d}$ and $ad-bc=1$, whence
\begin{equation}\label{calDexpl}
   {\cal D}g(z, \xi)=\Big(\frac{az+b}{cz+d},\,\frac{\xi}{(cz+d)^2}\Big).
\end{equation}
Given $(z, \xi)\in T^1\H^2$, the relation ${\cal D}g(i, i)=(z, \xi)$ 
has a unique solution $g\in {\rm PSL}(2, \R)$. This means that the induced mapping 

\begin{equation}\label{Upsidef}
   \Upsilon: T^1\H^2\to {\rm PSL}(2, \R),\quad (z, \xi)\mapsto g
\end{equation}
is a bijection. In general, $T^1\H^2$ and ${\rm PSL}(2, \R)$ are identified. 
\medskip

The following result will be needed later when consider reversibility. 

\begin{lemma}\label{LgDg} If $g=\Upsilon(z, \xi)$ and $g_1=\Upsilon(z, -\xi)$, 
then $g_1=gj$ for $j=[J]=\{-J, J\}$, where 
$J=\scriptsize \Big(\begin{array}{cc} 0 & 1 \\ -1 & 0\end{array}\Big)$ 
represents the M\"obius transform $z\mapsto -\frac{1}{z}$. 
\end{lemma}

\subsubsection{Decompositions of $\PSL(2, \R)$}

For $t\in\R$, denote  
\begin{eqnarray*}
   A_t & = & \bigg(\begin{array}{cc} e^{t/2} & 0 \\
   0 & e^{-t/2}\end{array}\bigg)\in\SL(2,\R), \quad  
   B_t=\bigg(\begin{array}{cc} 1 & t \\
   0 & 1\end{array}\bigg)\in \SL(2,\R), 
   \\ C_t & = & \bigg(\begin{array}{cc} 1 & 0 \\
      t & 1\end{array}\bigg)\in\SL(2,\R),\quad 
   D_t=\bigg(\begin{array}{cc} \cos(t/2) & \sin(t/2) \\
      -\sin(t/2) & \cos(t/2)\end{array}\bigg)\in\SL(2,\R)
\end{eqnarray*} 
and, respectively, 
\begin{equation}\label{abcdt} 
   a_t=[A_t]\in\G,\quad b_t=[B_t]\in\G,\quad 
   c_t=[C_t]\in\G,\quad d_t=[D_t]\in\G,
\end{equation} 
where $\G={\rm PSL}(2, \R)$. 
\medskip

Next we establish some useful factorizations for elements of ${\PSL}(2, \R)$. 
These results will be of central importance later when it comes 
to splitting the tangent space according to the hyperbolicity; 
recall that $\{gb_t: t\in\R\}$ and $\{gc_t: t\in\R\}$, respectively, 
give rise to the stable and unstable manifolds at $g\in {\rm PSL}(2, \R)$. 
First we consider the so-called `NAK decomposition' of $\SL(2,\R)$.

\begin{lemma}[\cite{Sieber1}]\label{lemnak}
Let $g=[G]\in\PSL(2,\R)$ for $G= \scriptsize\Big(\begin{array}{cc} a&b\\c&d\end{array}\Big)\in\SL(2,\R)$. 
Then $g= b_x a_{\ln y}d_\theta$ for 
\[ x=\frac{ac+bd}{c^2+d^2},\ \  y=\frac{1}{c^2+d^2},\ \  \theta=-2\arg(d+ic). \] 
\end{lemma}

\begin{lemma}\label{decompo}
Let $g=[G]\in {\rm PSL}(2, \R)$ for $G=\scriptsize \Big(\begin{array}{cc} a & b \\ c & d\end{array}\Big)
\in {\rm SL}(2, \R)$. 
\begin{itemize}
\item[(a)] If $a\neq 0$, then $g=c_u b_s a_t$ for 
\begin{equation}\label{ust-def}
   t=2\ln |a|,\quad s=ab,\quad u=\frac{c}{a}. 
\end{equation} 
\item[(b)] If $d\neq 0$, then $g=b_s c_u a_t$ for 
\[ t=-2\ln |d|,\quad s=\frac{b}{d},\quad u=cd. \] 
\end{itemize} 
\end{lemma} 
\noindent
{\bf Proof\,:} (a) Let $(t, s, u)$ be given by (\ref{ust-def}). To begin with, 
\[ C_u B_sA_t=\Bigg(\begin{array}{cc} e^{t/2} & se^{-t/2} \\ ue^{t/2} & (1+su)e^{-t/2}\end{array}\Bigg). \] 
If $a>0$, then $e^{t/2}=a$, $se^{-t/2}=b$, $ue^{t/2}=c$, and $(1+su)e^{-t/2}=(1+bc)/a=d$, 
using that $ad-bc=1$. Thus $C_u B_sA_t=G$ and $c_u b_sa_t=g$. 
If $a<0$, then $e^{t/2}=-a$, $se^{-t/2}=-b$, $ue^{t/2}=-c$, and $(1+su)e^{-t/2}=-(1+bc)/a=-d$, 
and hence $C_u B_sA_t=-G$ which yields once again that $a_t c_u b_s=g$. 
(b) Here the argument is analogous. {\hfill$\Box$}\bigskip

\subsubsection{Distance on $\PSL(2,\R)$}

There is a natural Riemannian metric on $\G=\PSL(2,\R)$ such that
the induced metric function $d_{\G}$ is left-invariant under
$\PSL(2,\R)$ and right-invariant under ${\rm PSO}(2)=\{d_t : t\in\R\}$, i.e., for $g_1, g_2 \in\PSL(2,\R)$: 
\[d_{\G}(g_1,g_2)=d_{\G}(hg_1,hg_2)=d_{\G}(g_1g,g_2g)\]
for all $h\in\PSL(2,\R)$ and $g\in {\rm PSO}(2)$. Let $e=[E_2]\in {\rm PSL}(2, \R)$ denote the neutral element. 

\begin{lemma}\label{distancePSL}
For $t, u, s, \theta\in\R$, we have
\begin{itemize}
\item[(a)] $ 
   d_{\G}(a_t,e)=\frac{|t|}{\sqrt 2}, \  
    d_{\G}(d_\theta,e)=\frac{|\theta|}{\sqrt 2},
    \ 
   d_{\G}(b_s, e)\le |s|, \ 
   d_{\G}(c_u, e)\le |u|.$
\item[(b)] $
   d_{\G}(a_{-t}b_sa_t,e)\leq |s|e^{-t},
 \   d_{\G}(a_{-t}c_ua_t,e)\leq |u| e^t.
$
 \item[(c)] $d_\G(b_s a_t d_\theta,e)\leq 
 \frac{1}{\sqrt 2}|t| +\frac{1}{\sqrt 2}|\theta|+|s|$.
\end{itemize}
\end{lemma}

\subsubsection{Traces and hyperbolic elements}

By means of the trace there is an important classification 
of the elements of ${\rm PSL}(2, \R)$. 

\begin{definition}[Trace] The trace of $g\in {\rm PSL}(2, \R)$ is 
${\rm tr}(g)=|{\rm tr}(A)|$, where $g=[A]=\{-A, A\}$. 
\end{definition} 

Observe that ${\rm tr}(-A)=-{\rm tr}(A)$ for a matrix $A\in {\rm SL}(2, \R)$, 
so the assignment ${\rm tr}(g)=|{\rm tr}(A)|$ makes sense. 
 
\begin{definition}[Elliptic, parabolic, hyperbolic element]
An element $g\in {\rm PSL}(2, \R)$ is said to be elliptic, if ${\rm tr}(g)<2$, 
parabolic, if ${\rm tr}(g)=2$, and hyperbolic, if ${\rm tr}(g)>2$. 
\end{definition} 


\subsection{The geodesic flow on $\Gamma\backslash \H^2$}

Throughout this paper we let $\Gamma\subset {\rm PSL}(2, \R)$ denote a Fuchsian group, 
which means that it is a discrete subgroup of $\PSL(2,\R)$. 

\subsubsection{$\Gamma\backslash\H^2$}
\label{gammodH2}

The Fuchsian group $\Gamma\subset {\rm PSL}(2, \R)$ induces a left action on $\H^2$ by the assignment
\[ \rho: \Gamma\times\H^2\to\H^2,
   \quad\rho(\gamma, z)=\Phi(\gamma)(z). \] 
Applying the identification of $\gamma$ with the M\"obius transform it generates, often this is written as $\rho(\gamma,z)=\gamma z$. 
The system of the $\Gamma$-orbits is
\[ \Gamma\backslash\H^2=\{\Gamma z: z\in\H^2\}, \]
where $\Gamma z=\{\gamma z: \gamma\in\Gamma\}$
denotes the orbit of $z\in\H^2$ under the left action $\rho$.
We denote by $\pi_\Gamma$ the natural projection 
\begin{equation}\label{pigam} 
   \pi_\Gamma:\H^2\rightarrow\Gamma\backslash \H^2,\quad  \pi_\Gamma(z)=[z]=\Gamma z. 
\end{equation}

\begin{theorem}
If the action $\rho$ of $\Gamma$ on $\H^2$ is
free (of fixed points) then 
$\Gamma\backslash\H^2$ has a Riemann surface structure.
\end{theorem}
\noindent
The genus of the resulting Riemann surface is at least two and it has the hyperbolic plane $\H^2$ 
as the universal covering. Furthermore, $\pi_\Gamma: \H^2\rightarrow \Gamma\backslash\H ^2$ 
from (\ref{pigam}) becomes a local isometry. This implies that $\Gamma\backslash\H^2$ also has curvature $-1$, 
and the geodesics on $\Gamma\backslash\H^2$ are the projections of the geodesics on $\H^2$. 
\medskip 

Denote by $(g_{\,\Gamma,p})_{p\,\in\,\Gamma\backslash\H^2}$ the natural Riemannian metric on $\Gamma\backslash\H^2$. 
Its unit tangent bundle is
\[ T^1(\Gamma\backslash\H^2)=\big\{(p, \xi): p\in\Gamma\backslash\H^2,
   \,\xi\in T_p(\Gamma\backslash\H^2),
   \,{\|\xi\|}_p=g_{\,\Gamma,\,p}(\xi, \xi)^{1/2}=1\big\}. \]
Let ${(\varphi_t^\X)}_{t\in\R}$ be the geodesic flow on $\X=T^1(\Gamma\backslash\H^2)$. 


\subsubsection{$\Gamma\backslash {\rm PSL}(2, \R)$}

Instead of considering $\X=T^1(\Gamma\backslash\H^2)$ it is often easier to work on $X=\Gamma\backslash {\rm PSL}(2, \R)$, 
which is the system of right co-sets $\Gamma g=\{\gamma g: \gamma\in\Gamma\}$.
In general $\Gamma\backslash {\rm PSL}(2, \R)$ will not be a group;
see \cite[example 9.15]{einsward}. The associated canonical projection is 
\begin{equation}\label{ascapr} 
   \Pi_\Gamma: {\rm PSL}(2, \R)\to\Gamma\backslash {\rm PSL}(2, \R),
   \quad\Pi_\Gamma(g)=[g]=\Gamma g.
\end{equation} 
Since $\Pi_\Gamma$ is continuous and onto, and since ${\rm PSL}(2, \R)$ 
is connected, also $\Gamma\backslash {\rm PSL}(2, \R)$ is connected. 
Furthermore, $X=\Gamma\backslash {\rm PSL}(2, \R)$
is a three-dimensional real analytic manifold.
We define a flow ${(\varphi_t^X)}_{t\in\R}$ on $X$ by
\[ \varphi_t^X(x)=\Gamma(g a_t),
   \quad t\in\R,\quad x=\Gamma g\in X. \]  
The flow ${(\varphi_t^X)}_{t\in\R}$ is smooth and satisfies $\Pi_\Gamma\circ\varphi_t^\G
=\varphi_t^X\circ\Pi_\Gamma$, where $\Pi_\Gamma: \G\to X$ denotes the natural projection from (\ref{ascapr}) above 
and $\varphi_t^\G(g)=ga_t$ is a flow on $\G=\PSL(2,\R)$. This shows how the flow $\varphi_t^\G$ on $\G$ induces 
the `quotient flow' $\varphi_t^X$ on $X$.
\medskip   
   
The following statement makes rigorous the identification 
$T^1(\Gamma\backslash\H^2)\cong\Gamma\backslash {\rm PSL}(2, \R)$.

\begin{theorem}\label{Xidef} Let $\Gamma\subset {\rm PSL}(2, \R)$ be a Fuchsian group
such that the left action of $\Gamma$ on $\H^2$ is free. Then there is a bijection
\[ \Xi:\,\,T^1(\Gamma\backslash\H^2)\to\Gamma\backslash {\rm PSL}(2, \R). \]
With the Sasaki metric (see \cite{pater}) on $T^1(\Gamma\backslash\H^2)$ 
induced by the natural Riemannian metric on $\Gamma\backslash\H^2$ 
and the natural Riemannian metric on $\Gamma\backslash\PSL(2,\R)$, 
$\Xi$ is an isometry. Furthermore, we have the conjugation relation 
\begin{equation}\label{geodflcalX} 
   \varphi_t^{{\cal X}}=\Xi^{-1}\circ\varphi_t^X\circ\Xi
\end{equation} 
for $t\in\R$.
\end{theorem}
\noindent
Thus instead of the geodesic flow ${(\varphi_t^\X)}_{t\in\R}$ on $\X$ 
from section \ref{gammodH2} we can study the flow ${(\varphi_t^X)}_{t\in\R}$ on $X$.  
This viewpoint offers several advantages. For instance, 
one can calculate explicitly the stable and unstable manifolds at a point $x\in X$ to be
\[ W^s_X(x)=\{\theta^X_t(x),t\in\R\}
   \quad \mbox{and}\quad W^u_X(x)=\{\eta^X_t(x), t\in\R\}, \]
where ${(\theta^X_t)}_{t\in\R}$ and ${(\eta^X_t)}_{t\in\R}$ are the horocycle flow 
and conjugate horocycle flow given by 
\[ \theta^X_t(\Gamma g)=\Gamma (g b_t)\quad\mbox{and}\quad\eta^X_t(\Gamma g)=\Gamma (g c_t); \] 
recall $b_t$ and $c_t$ from (\ref{abcdt}). Furthermore, ${(\varphi^X_t)}_{t\in\R}$ is an Anosov flow: 
For every $x\in X$ there exists a splitting of the tangent space
\[ T_x X= E^0(x)\oplus E^s(x)\oplus E^u(x) \]
such that ${(d\varphi_t^X)}_x(E^s(x))=E^s(\varphi_t^X(x))$ 
and ${(d\varphi_t^X)}_x(E^u(x))=E^u(\varphi_t^X(x))$ for $t\in\R$ 
as well as $\|{{(d\varphi_t^X)}_x|}_{E^s(x)}\|\le C_1\,e^{-\lambda_1 t}$ 
and $\|{{(d\varphi_{-t}^X)}_x|}_{E^u(x)}\|\le C_2\,e^{-\lambda_2 t}$ 
for $t\in [0, \infty[$; here $C_1, C_2>0$ and $\lambda_1, \lambda_2>0$ 
are independent of $x$. One can take  $E^0(x)={\rm span}\{\frac{d}{dt}\,\varphi_t^X(x)|_{t=0}\}$ and 
\[ E^s(x)={\rm span}\Big\{\frac{d}{dt}\,\theta_t^X(x)\Big|_{t=0}\Big\},
   \quad E^u(x)={\rm span}\Big\{\frac{d}{dt}\,\eta_t^X(x)\Big|_{t=0}\Big\}. \] 

The flow ${(\varphi_t^X)}_{t\in\R}$ 
also enjoys a useful reversibility property. 

\begin{lemma}\label{revpropX} 
If $\varphi_t^X(x)=y$ for some $t\in\R$, $x=(\Pi_\Gamma\circ\Upsilon)(z, \xi)\in X$, 
and $y=(\Pi_\Gamma\circ\Upsilon)(z, \zeta)\in X$, then $\varphi_t^X(y')=x'$ 
for $x'=(\Pi_\Gamma\circ\Upsilon)(z, -\xi)\in X$ and $y'=(\Pi_\Gamma\circ\Upsilon)(z, -\zeta)\in X$. 
\end{lemma} 
\noindent
{\bf Proof\,:} Denote $g=\Upsilon(z, \xi)\in \G$, $h=\Upsilon(z, \zeta)\in \G$, 
$g'=\Upsilon(z, -\xi)\in \G$, and $h'=\Upsilon(z, -\zeta)\in \G$. 
According to lemma \ref{LgDg} we have $g'=gj$ and $h'=hj$ for $j=[J]=\{-J, J\}$, 
where $J=\scriptsize \Big(\begin{array}{cc} 0 & 1 \\ -1 & 0\end{array}\Big)$. 
Now $\varphi_t^X(x)=y$ reads as $\Pi_\Gamma(ga_t)=\Pi_\Gamma(h)$ 
or $ga_t=\gamma h$ for some $\gamma\in\Gamma$. Then $j a_t j^{-1}=a_{-t}$ yields 
$h' a_t=h' j^{-1} a_{-t} j=ha_{-t} g^{-1} g'=(ga_t h^{-1})^{-1} g'=\gamma^{-1} g'$, 
so that $\Pi_\Gamma(h' a_t)=\Pi_\Gamma(g')$ and hence $\varphi_t^X(y')=x'$. 
{\hfill$\Box$}\bigskip

We define a metric function $d_{\Gamma\backslash\G}$ on $X=\Gamma\backslash\G$ by 
\[ d_{\Gamma\backslash \G}(x_1, x_2)
   =\inf_{\gamma_1, \gamma_2\in\Gamma} d_{\G}(\gamma_1 g_1, \gamma_2 g_2)
   =\inf_{\gamma\in\Gamma} d_{\G}(g_1, \gamma g_2), \]   
where $x_1=\Pi_\Gamma(g_1)$, $x_2=\Pi_\Gamma(g_2)$,
and $d_{\G}$ denotes the metric function on $\G$. 
In fact, if $\Gamma\backslash\G$ is compact, one can prove that the infimum is a minimum:
\begin{equation}
d_{\Gamma\backslash \G}(x_1, x_2)=\min_{\gamma\in\Gamma} d_{\G}(g_1, \gamma g_2).
\end{equation} 
It is possible to derive 
a uniform lower bound on $d_{\G}(g, \gamma g)$ 
for $g\in \G$ and $\gamma\in\Gamma\setminus\{e\}$, 
provided that $X=\Gamma\backslash \G$ is compact. 

\begin{lemma}\label{injrad2} 
If $X=\Gamma\backslash \G$ is compact, then there is $\sigma_0>0$ such that 
$d_{\G}(g, \gamma g)\ge\sigma_0$ for all $g\in \G$ and $\gamma\in\Gamma\setminus\{e\}$.
\end{lemma} 
\medskip 

Compactness of the quotients can be characterized as follows. 

\begin{theorem} Let the left action of $\Gamma$ on $\H^2$ be free. 
Then there are equivalent: 
\begin{itemize}
\item[(a)] $\Gamma\backslash\H^2$ is compact.
\item[(b)] $T^1(\Gamma\backslash\H^2)$ is compact. 
\item[(b)] $\Gamma\backslash {\rm PSL}(2, \R)$ is compact.
\item[(c)] $\overline{D_{z_0}(\Gamma)}\subset\H^2$ is compact for every $z_0\in\H^2$ 
such that $z_0\neq\gamma z_0$ for all $\gamma\in\Gamma\setminus\{e\}$; 
here $D_{z_0}(\Gamma)$ denotes the Dirichlet region fundamental domain (see \cite[definition 11.3]{einsward}).  
\end{itemize}
In this case there is $\eps_0>0$ such that ${\rm tr}(\gamma)\ge 2+\eps_0$ 
holds for all $\gamma\in\Gamma\setminus\{e\}$, and in particular every $\gamma\in\Gamma\setminus\{e\}$ 
is hyperbolic.  
\end{theorem} 
\medskip

\begin{definition}[Periodic point]
A point $x\in X$ is called periodic under the flow ${(\varphi_t^X)}_{t\in\R}$ on $X$, 
if there is $T\in\R$ such that
\begin{equation}\label{pedf}
   \varphi_T^X(x)=x.
\end{equation}
The smallest $T>0$ satisfying (\ref{pedf}) is called the (minimal) period.  
\end{definition}
Periodic points (or orbits) of the geodesic flow ${(\varphi_t^\X)}_{t\in\R}$ on $\X$ are defined similarly. 

\begin{lemma}\label{exgeodper} Suppose that every element in $\Gamma\setminus\{e\}$ 
is hyperbolic. Then for every $\gamma\in\Gamma\setminus\{e\}$ and $T>0$ such that 
\[ e^{T/2}+e^{-T/2}={\rm tr}(\gamma), \]  
there exists $g\in \G$ so that $\gamma=g a_T g^{-1}$. 
In particular, $x=\Pi_\Gamma(g)\in X$ is a $T$-periodic point 
of the flow ${(\varphi_t^X)}_{t\in\R}$ on $X$.
\end{lemma} 

\begin{remark}{\rm There exist bijections 
\[ {{\cal PO}}_X\longleftrightarrow {\cal CG}_Y\longleftrightarrow\mathfrak{C}_\Gamma \] 
between the class of periodic orbits ${{\cal PO}}_X$ 
of the flow ${(\varphi_t^X)}_{t\in\R}$ on $X=\Gamma\backslash {\rm PSL}(2, \R)$, 
the class ${\cal CG}_Y$ of oriented unit speed closed geodesics on $Y=\Gamma\backslash\H^2$, 
and the system $\mathfrak{C}_\Gamma$ of conjugacy classes of primitive elements in $\Gamma\backslash\{e\}$; 
here the conjugacy class of $\gamma\in\Gamma$ is ${\{\gamma\}}_\Gamma=\{\sigma\gamma\sigma^{-1}: \sigma\in\Gamma\}$ 
and an element $\gamma\in\Gamma$ is called primitive, if $\gamma=\eta^m$ for some $\eta\in\Gamma$ 
and $m\in\Z$ implies that $m=1$ or $m=-1$.
Furthermore, if $T>0$ denotes the prime period of $c\in {{\cal PO}}_X$ 
and ${\{\gamma\}}_\Gamma\leftrightarrow c$, then ${\rm tr}(\gamma)=e^{T/2}+e^{-T/2}$. 
Next, if $s\in {\cal CG}_Y$ satisfies $s\leftrightarrow {\{\gamma\}}_\Gamma$ 
and $\tau>0$ denotes the prime period of $s$, then the curve length $\ell(s)$ of $s$ 
is $\tau$, since $s$ is a unit speed geodesic. Also ${\rm tr}(\gamma)=e^{\tau/2}+e^{-\tau/2}$. 
This yields the relation $T=\tau=\ell(s)$ between the period of $c$ and the curve length of $s$. 
Finally, by definition, the length of $\gamma$ is $\delta(S)=\inf_{z\in\H^2} d_{\H^2}(z, S(z))$, 
where $S=\Phi(\gamma)$. Then one can show that ${\rm tr}(\gamma)=2\cosh(\delta(S)/2)$, 
and hence $\delta(S)=T=\tau=\ell(s)$ is found, which also explains why $\delta(S)$ is called a length. 
{\hfill$\diamondsuit$}
}
\end{remark}
\medskip

We shall also need local stable and unstable manifolds. 

\begin{definition}[Local stable and unstable manifolds]\label{locsu-mfk}
Let $\eps>0$ and $x\in X$. Then 
\[ W^s_{X,\,\eps}(x)=\{\theta_t^X(x): |t|<\eps\}=\{\Gamma(gb_t): |t|<\eps\} \]
and
\[ W^u_{X,\,\eps}(x)=\{\eta_t^X(x): |t|<\eps\}=\{\Gamma(gc_t): |t|<\eps\} \] 		 
for $x=\Gamma g$ are called the {\em local stable} and {\em local unstable manifold} 
of $x$ of size $\eps$, respectively. 
\end{definition} 

Note that both sets do not depend on the choice of $g\in \G$ such that $x=\Gamma g$. 
The following definition is modified from \cite{bieder}. 

\begin{figure}[ht]
\begin{center}
\begin{minipage}{0.6\linewidth}
   \centering
   \includegraphics[angle=0,width=0.7\linewidth]{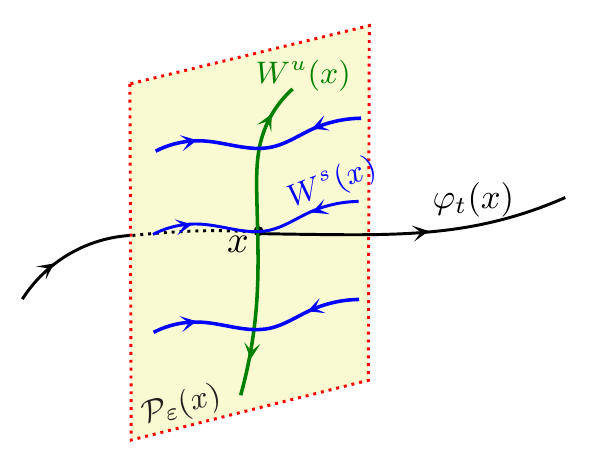}
\end{minipage}
\end{center}
\caption{Poincar\'e section}\label{pcsp}
\end{figure}

\begin{definition}[Poincar\'e section]
Let $x\in X$ and $\eps>0$. The \textit{Poincar\'e section} of radius $\eps$ at $x$ is 
\begin{equation*}
  \P_\eps(x)=\{(\theta^X_s\circ\eta^X_u)(x): |s|<\eps, |u|<\eps\}=\{\Gamma(g c_ub_s) : |u|<\eps, |s| <\eps\},
\end{equation*}
where $g\in \G$ is such that $x=\Gamma g$ (see Figure \ref{pcsp}). 
\end{definition}
\noindent
Notice that $\P_\eps(x)$ is independent of $g$, since in the case where $x=\Gamma g_1=\Gamma g_2$ 
we have $\gamma g_1=g_2$ for some $\gamma\in\Gamma$. Hence $\gamma g_1c_ub_s=g_2c_ub_s$,
and consequently $\Gamma(g_1c_ub_s)=\Gamma(g_2c_ub_s)$.

\subsection{Some estimates}

In this subsection we collect some useful technical results. 

\begin{lemma}\label{konvexa}
(a) For every $\eps>0$ there is $\delta>0$ with the following property. If $G\in\SL(2,\R)$ satisfies
$\|G-E_2\|<\delta$ then $d_{{\rm PSL}(2, \R)}(g, e)<\eps$ for $g=\pi(G)$, 
where $\pi: {\rm SL}(2, \R)\to {\rm PSL}(2, \R)$ is the natural projection.  

\smallskip
\noindent
(b) For every $\eps>0$ there is $\delta>0$ with the following property. 
If $d_{\PSL(2,\R)}(g,h)<\delta$ then there are
\[ G=\Bigg(\begin{array}{cc}g_{11}&g_{12}\\
   g_{21}&g_{22}
   \end{array}\Bigg)\quad \mbox{and}\quad 
   H=\Bigg(\begin{array}{cc}h_{11}&h_{12}\\
   h_{21}&h_{22}\end{array}\Bigg) \]
such that $g=\pi(G), h=\pi(H)$ and
$|g_{11}-h_{11}|+|g_{12}-h_{12}|+|g_{21}-h_{21}|+|g_{22}-h_{22}|<\eps$.
\end{lemma}
\noindent
{\bf Proof\,:}
(a)
Fixing $\eps>0$, we define $\delta=\min\big\{\frac 18,\frac{\eps}{13}\big\}$. 
Let $g=\pi(G)\in\PSL(2,\R)$ for $G=\scriptsize\Big(\begin{array}{cc} a&b\\c&d\end{array}\Big)\in\SL(2,\R)$ be such that  
\[\|G-E_2\|=|a-1|+|b|+|c|+|d-1|<\delta. 
\]
Then we have
\begin{equation}\label{abc} 
   |a|<1+\delta,\quad |b|<\delta,\quad |c|<\delta,\quad 0<d<1+\delta.
\end{equation} 
Use lemma \ref{lemnak} to write $g=b_sa_td_\theta$ for  
\[s=\frac{ac+bd}{c^2+d^2}, \quad t=-\ln(c^2+d^2),\quad
\theta=-2\arg(d+ic)=-2\,{\rm arctan}(c/d),\]
noting that $d>0$. 
Applying lemma \ref{distancePSL}(c), we obtain
\begin{eqnarray*}
   d_\G(g, e) & \leq & \Big|\frac{ac+bd}{c^2+d^2}\Big|
   +|\ln(c^2+d^2)|+2\,|\,{\rm arctan}(c/d)|
   \\ & \leq & 2\,|ac+bd| +2\,|c^2+d^2-1|+2\,\Big|\frac{c}d\Big|
   \\ & < & 4\,(\delta+1)\delta +2\,(\delta^2+2\delta) +4\delta
   < 13\delta\leq \eps,
\end{eqnarray*}
using (\ref{abc}), $c^2+d^2>\frac12$ and $|\ln(1+x)|<2|x|$ for $|x|<\frac12$.

\smallskip
\noindent
(b) Indeed, otherwise there are $\eps_0>0$ and $h=\pi(H)$ such that
\begin{equation}\label{gj-}
   |g_{11}^j-h_{11}|+|g_{12}^j-h_{12}|
   +|g_{21}^j-h_{21}|+|g_{22}^j-h_{22}|\geq \eps_0
\end{equation}
for some sequence $d_{\PSL(2,\R)}(g^j,h)\rightarrow 0$
and for all $G^j\in\SL(2,\R)$ such that $g^j=\pi(G^j)$. 
By the left-invariance of $d_{\PSL(2,\R)}$,
we have $d_{\PSL(2,\R)}(h^{-1}g^j,e)\rightarrow 0$. 
For $j\in\N$, take any $G^j$ such that $g^j=\pi(G^j)$. From (a) we deduce that
\[ |g_{22}^jh_{12}-g_{12}^jh_{22}|+|-g_{21}^jh_{11}+g_{22}^jh_{21}|\rightarrow 0,
   \ |g_{22}^jh_{11}-g_{12}^jh_{21}|\rightarrow 1,\ |-g_{21}^jh_{12}+g_{11}^jh_{22}|\rightarrow 1, \] 
and \[(g_{22}^jh_{12}-g_{12}^jh_{22})(-g_{21}^jh_{12}+g_{11}^jh_{22})\rightarrow 1.\]
Thus, along a subsequence which is not relabeled, either 
\[ -g_{21}^jh_{12}+g_{11}^jh_{22}\rightarrow 1,
   \ -g_{21}^jh_{12}+g_{11}^jh_{22}\rightarrow 1 \]
or 
\[-g_{21}^jh_{12}+g_{11}^jh_{22}\rightarrow -1,\ -g_{21}^jh_{12}+g_{11}^jh_{22}\rightarrow -1. \]  
The first case yields that 
$g_{11}^j\rightarrow -h_{11}, \,g_{12}\rightarrow -h_{12},\,
g_{21}^j\rightarrow -h_{21}$, and $g_{22}^j\rightarrow -h_{22}$.
We consider $\tilde G_j=-G_j$ which also has $g^j=\pi(\tilde G^j)$. 
But then (\ref{gj-}) implies
\[ |\tilde g_{11}^j+h_{11}| +|\tilde g_{12}^j+h_{12}|+|\tilde g_{21}^j+h_{21}|+|\tilde g_{22}^j+h_{22}|\geq \eps_0 \] 
which is impossible. In the second case we have
$g_{11}^j\rightarrow h_{11},\,g_{12}^j\rightarrow h_{12},\,g_{21}^j\rightarrow h_{21}$, 
and $g_{22}^j\rightarrow h_{22}$ and once more this is impossible.
{\hfill$\Box$}\bigskip

\begin{lemma}\label{lmexistsigma}
If $|u|,\, |s|<\frac14$ and $T\ge 1$, then the equation 
\begin{equation}\label{imgs} 
   -se^T\sigma^2+((1+su)e^T-1)\sigma+u=0
\end{equation} 
has a solution $\sigma\in\R$ such that $|\sigma|< 2|u|e^{-T}$. 
\end{lemma}
\noindent 
{\bf Proof\,:} 
To establish this assertion, consider first the case where $s=0$. Then 
$\sigma=\frac{u}{1-e^T}$ is the solution, and moreover $|\sigma|=\frac{|u|e^{-T}}{|1-e^{-T}|}
\leq \frac{|u|e^{-T}}{1-e^{-1}}< 2|u|e^{-T}$. For $s\neq 0$ consider the discriminant 
\begin{eqnarray}\label{hapin} 
   \Delta & = & ((1+su)e^T-1)^2+4sue^T
  = (1-su e^T)^2+e^{2T}(1+2su-2e^{-T}+4sue^{-T}) 
   \nonumber
   \\ & \ge & (1-su e^T)^2+e^{2T}\Big(1-2\cdot\frac{1}{4}\cdot\frac{1}{4}-2\cdot\frac{3}{8}
   -4\cdot\frac{1}{4}\cdot\frac{1}{4}\cdot\frac{3}{8}\Big)
   \nonumber
   \\ & = & (1-su e^T)^2+\frac{e^{2T}}{32}>(1-su e^T)^2
\end{eqnarray} 
and take the solution 
\[ \sigma=\frac{(1+su)e^T-1-\sqrt{\Delta}}{2se^T} \] 
to (\ref{imgs}). Then $1-sue^T\le |1-sue^T|\le\sqrt{\Delta}$ by (\ref{hapin}), 
so that 
\begin{eqnarray*} 
   |\sigma|
   =\frac{2|u|}{e^T-(1-su e^T)+\sqrt{\Delta}}
    \le  2|u|e^{-T}, 
\end{eqnarray*} 
as was to be shown. 
{\hfill$\Box$}\bigskip
  
\setcounter{equation}{0}  
  
\section{Sieber-Richter pairs}
\label{srp_sect} 

In this section we establish the existence of a partner orbit for a given periodic orbit
with small-angle self-crossing in configuration space. If the crossing angle is small enough 
and the surface is compact, then the partner is unique. The partner avoids crossing
in the encounter region and has a smaller period
as compared to the original one. We also derive an estimate for the action difference between them.

\subsection{Shadowing lemma and Anosov closing lemma}

For brevity we write $\G={\rm PSL}(2, \R)$, $X=\Gamma\backslash {\rm PSL}(2, \R)
=\Gamma\backslash \G$, and ${(\varphi_t)}_{t\in\R}={(\varphi_t^X)}_{t\in\R}$. 

\subsubsection{Shadowing lemma}

Recall definition \ref{locsu-mfk} of the local stable and unstable 
manifolds $W^s_{X,\,\eps}$ and $W^u_{X,\,\eps}$, 
and see \cite{bieder} for a similar result. 
\begin{figure}[ht]
\begin{center}
\begin{minipage}{0.8\linewidth}
   \centering
   \includegraphics[angle=0,width=0.7\linewidth]{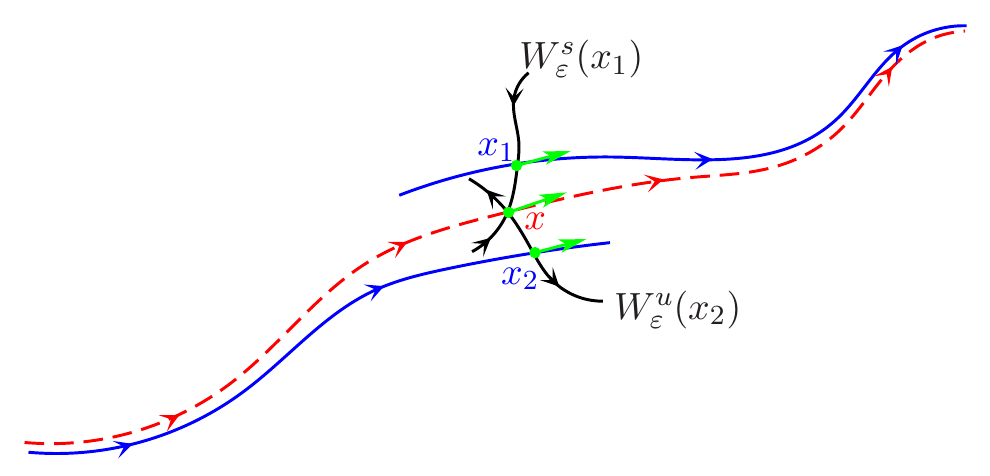}
\end{minipage}
\end{center}
\caption{Shadowing lemma}\label{sha}
\end{figure}

\begin{theorem}[Shadowing lemma]\label{shadlemII}  
If $\eps>0$, $x_1, x_2\in X$, and $x\in W^s_{X,\,\eps}(x_1)\cap W^u_{X,\,\eps}(x_2)$, then 
\[ d_X(\varphi_t(x_1), \varphi_t(x))<\eps e^{-t}
   \quad\mbox{for all}\quad t\in [0, \infty[ \]
and 
\[ d_X(\varphi_t(x_2), \varphi_t(x))<\eps e^t 
   \quad\mbox{for all}\quad t\in\, ]-\infty, 0]. \]
\end{theorem}
\noindent{\bf Proof\,:} Fix $g_1, g_2\in \G$ such that $\Pi_\Gamma(g_1)=x_1$ and $\Pi_\Gamma(g_2)=x_2$. 
By definition of $W^s_{X,\,\eps}(x_1)$ and $W^u_{X,\,\eps}(x_2)$ there are $|s|<\eps$ and $|u|<\eps$ 
so that $x=\Gamma(g_1 b_s)$ as well as $x=\Gamma(g_2 c_u)$. Then for $t\in [0, \infty[$ 
by lemma \ref{distancePSL}(b): 
\begin{eqnarray*} 
   d_X(\varphi_t(x_1), \varphi_t(x))
   & = & \inf_{\gamma\in\Gamma} d_{\G}(g_1 a_t, \gamma g_1 b_s a_t)
   \le d_{\G}(g_1 a_t, g_1 b_s a_t)
   = d_{\G}(a_t, b_s a_t)
   \\ &=&d_{\G}(a_{-t} b_{s} a_t, e)
   \le |s|e^{-t}<\eps e^{-t}. 
\end{eqnarray*}  
Similarly, for $t\in \,]-\infty, 0]$, 
\begin{eqnarray*} 
   d_X(\varphi_t(x_2), \varphi_t(x))
   & = & \inf_{\gamma\in\Gamma} d_{\G}(g_2 a_t, \gamma g_2 c_u a_t)
   \le d_{\G}(g_2 a_t, g_2 c_u a_t)
     =  d_{\G}(a_t, c_u a_t)
   \\
   &=& d_{\G}(a_{-t} c_u a_t, e)\le |u|e^t<\eps e^t, 
\end{eqnarray*}  
completing the proof. {\hfill$\Box$}\bigskip

\subsubsection{Anosov closing lemma}

There are many different versions of the Anosov closing lemma. 
In the following we shall need a very quantitative version. 
See \cite{bieder} and \cite[Lemma 7.4]{EinsDuke} for results in a similar vein. 
Let $\sigma_0>0$ be chosen according to lemma \ref{injrad2}: 
If $g\in \G$ and $\gamma\in\Gamma\setminus\{e\}$, then 
$d_G(g, \gamma g)\ge\sigma_0$ is verified. 

\begin{theorem}[Anosov closing lemma]\label{anosov1}
Suppose that $\eps\in\, ]0, \frac{1}{4}[$, 
$x\in X$, $T\ge 1$, and $\varphi_T(x)\in {\cal P}_\eps(x)$. 
Let $x=\Gamma g$ and  $\varphi_T(x)=\Gamma gc_ub_s$ 
for $g\in\PSL(2,\R), |u|<\eps,|s|<\eps$. 
Then there are $x'\in {\cal P}_{2\eps}(x)$ such that $x'=\Gamma gc_\sigma b_\eta$ 
and $T'\in\R$ so that 
\begin{equation}\label{x'}
   \varphi_{T'}(x')=x'\quad \mbox{and}\quad d_X(\varphi_t(x),\varphi_t(x'))
   <2|u|+|\eta|< 4\eps\quad \mbox{for all}\quad t\in[0,T].
\end{equation}
Furthermore,
\begin{equation}\label{T-T'1}
   \Big|\frac{T'-T}2-\ln(1+us)\Big| <5|us|e^{-T},
\end{equation}
\begin{equation}\label{TT'anosov1}
   e^{T'/2}+e^{- T'/2}=e^{T/2}+e^{-T/2}+us e^{T/2},
\end{equation}
and
\begin{eqnarray}
   |\sigma|<2|u|e^{-T},\quad |\eta-s|<2s^2|u|+2|s|e^{-T}.
\end{eqnarray}
\end{theorem}

\begin{figure}[ht]
\begin{center}
\begin{minipage}{0.55\linewidth}
   \centering
   \includegraphics[angle=0,width=0.7\linewidth]{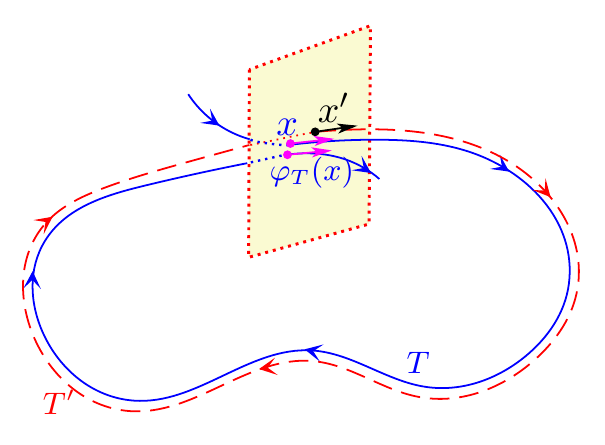}
\end{minipage}
\end{center}
\caption{Anosov closing lemma}\label{Anosovlemma}
\end{figure}

\noindent
{\bf Proof\,:} Write $x=\Pi_\Gamma(g)$ for $g\in \G$. Then 
$ \varphi_T(x)=\Pi_\Gamma(ga_T)= \Pi_\Gamma( gc_ub_s)$ and there is $\zeta\in\Gamma$ such that
$\zeta=gc_ub_sa_{-T}g^{-1}.$
First we apply lemma \ref{lmexistsigma} to obtain a solution $\sigma\in\R$ of the equation
\begin{equation*}
   -se^{T}\sigma^2+((1+su)e^{T}-1)\sigma+u=0
\end{equation*}
such that $|\sigma|<2|u|e^{-T}$. Define
\[ \eta=\frac{s}{1+s(u-\sigma)-s\sigma -e^{-T}}. \] 
Now $1+su-2s\sigma -e^{-T} >1-\frac1{16}-(1+\frac{1}{8})e^{-T}>\frac12$, 
so $\eta$ is well-defined and $|\eta|<2|s|<2\eps$. Then 
\[ |\eta-s|=\Big|\frac{s^2u-2s^2\sigma-se^{-T}}{1+su-2s\sigma-e^{-T}}\Big|\leq 2s^2|u|+2|s|e^{-T}, \] 
owing to $|\sigma|<2|u|e^{-T}$. Put $g'=g c_\sigma b_\eta\in \G$ and $x'=\Pi_\Gamma(g')$
to obtain $x'\in\P_{2\eps}(x)$. Define
\[ T'=T+2\ln(1+su-s\sigma).\]
We have $1+su-s\sigma>1-(\frac1{16}+\frac14\frac12)=\frac{13}{16}.$
Therefore ${T}'$ is well-defined and
\[ \frac{T'-T}2= \ln(1+su-s\sigma) =\ln(1+su)+\ln\Big(1-\frac{s\sigma}{1+su}\Big). \] 
Now we estimate the last term.
Note that $1-\frac{s\sigma}{1+su}>1-\frac{\frac{1}{16}}{1-\frac1{16}}=\frac{14}{15}$
and $|\ln(z)|<2|z-1|$ for $z\in [\frac{14}{15},\infty[$ yields
\[ \Big|\ln\Big(1-\frac{s\sigma}{1+su}\Big)\Big|<2\frac{|s\sigma|}{|1+su|}
   <2\frac{2|us|e^{-T}}{1-\frac16}<5|us|e^{-T}, \]
using $|\sigma|<2|u|e^{-T}$.
In addition, let $\tilde  g=gc_\sigma$ and $\tilde x=\Pi_\Gamma(\tilde g)$. For
\[ \hat g=\tilde g^{-1}\zeta\tilde g a_T=c_{-\sigma}g^{-1}\zeta g c_\sigma a_T
   =c_{\sigma}g^{-1}gc_ub_sa_{-T}g^{-1}g c_\sigma a_T
   =c_{u-\sigma}b_sc_{\sigma e^{T}} \]
observe that the matrix
\begin{eqnarray*}
   \hat A & = & C_{u-\sigma}B_sC_{\sigma e^{T}}
   =\Bigg(\begin{array}{cc}1+s\sigma e^{T}& s \\ -se^{T}\sigma^2+((1+su)e^{T}-1)\eta+u&1+s(u-\sigma)
   \end{array}\Bigg)\\
   & = & \Bigg(\begin{array}{cc}1+s\sigma e^{T}&s\\ 0& 1+s(u-\sigma)\end{array}\Bigg)
   =\Bigg(\begin{array}{cc}\hat a_{11} &\hat a_{12}\\ 0& \hat a_{22}\end{array}\Bigg)\in\SL(2,\R)
\end{eqnarray*}
satisfies $\pi(\hat A)=\hat g$. Furthermore,
\begin{eqnarray*}
   {B_{-\eta}\hat A A_{-T} B_\eta A_{T'}}
   =\Bigg(\begin{array}{cc} e^{(T'-T)/2}\hat a_{11} 
   & \eta(e^{-(T+T')/2}\hat a_{11} -e^{(T-T')/2}\hat a_{22})+\hat a_{12} e^{(T-T')/2} 
   \\0 & e^{(T-T')/2}\hat a_{22}\end{array}\Bigg).
\end{eqnarray*}
Due to $T'=T+2\ln\hat a_{22}$ and $\hat a_{11}\hat a_{22}=1$ we see that
$e^{(T-T')/2}\hat a_{22}=1$ as well as $e^{(T'-T)/2}\hat a_{11}=1$.
Then 
\[ \eta=\frac{s}{1+(u-\sigma)s-\sigma s-e^{-T}}
   =\frac{\hat a_{12}}{\hat a_{22}-\hat a_{11}e^{-T}} \]
and consequently also
\[ \eta(e^{-(T+T')/2}\hat a_{11} -e^{(T-T')/2}\hat a_{22})+\hat a_{12} e^{(T-T')/2}=0. \] 
In summary, we have shown that
$B_{-\eta}\hat A A_{-T} B_\eta A_{T'}=E_2$, and thus
\begin{equation}\label{c}
   b_{-\eta}\hat g a_{-T}b_\eta a_{T'}=e.
\end{equation}
Now we are in a position to verify (\ref{x'}). From (\ref{c}) and the various definition
\[ \zeta g c_\sigma b_\eta a_{T'}
   =\zeta(\zeta^{-1}gc_\sigma \hat g a_{-T})b_\eta a_{T'}
   =g c_\sigma\hat g a_{-T}b_\eta a_{T'}
   =gc_\sigma b_\eta .\]
Since $\zeta\in\Gamma$, this implies that
\[\varphi_{T'}(x')=\Gamma g'a_{T'}=\Gamma g c_\sigma b_\eta a_{T'}=\Gamma g c_\sigma b_\eta=\Gamma g'=x'.\]
Finally, for $t\in[0,T]$ by the left-invariance of $d_{\G}$
and lemma \ref{distancePSL}(b)
\begin{eqnarray*}
   d_X(\varphi_t(x),\varphi_t(x'))&\leq& d_X(\varphi_t(x),\varphi_t(\tilde x))+d_X(\varphi_t(\tilde x),\varphi_t(x'))\\
   & \leq & d_{\G}(ga_t, g c_\sigma a_t)+d_{\G}(gc_\sigma a_t,g  c_\sigma b_\eta a_t)\\
   & = & d_{\G}(a_{-t} c_{-\sigma} a_t,e)+d_{\G}(a_{-t}b_{-\eta}a_{t},e)\\
   & \leq & |\sigma|e^{t}+|\eta|e^{-t}\leq 2|u|e^{t-T}+2|s|e^{-t}\\
   & < & 2|u|+|\eta|< 4\eps.
\end{eqnarray*}
Due to $\zeta=gc_ub_sa_{-T}g^{-1}$, this implies $\tr(\zeta)=\tr(c_ub_sa_{-T})=e^{T/2}+e^{-T/2}+sue^{-T/2}$ 
and we obtain (\ref{TT'anosov1}).
{\hfill$\Box$}\bigskip

\begin{remark}\rm 
In the setting of the Anosov closing lemma: 
\smallskip

\noindent
(a) From (\ref{TT'anosov1}) we see that the period $T'$ of the partner is determined by
\begin{equation}\label{formT'}
T'=2\,\arccosh\Big(\frac{e^{T/2}+e^{-T/2}+us e^{T/2}}2\Big).
\end{equation} 
\noindent
(b) By (\ref{T-T'1}): $|T-T'|<2|\ln(1+us)|+10|us|e^{-T}<4\eps^2+10\eps^2e^{-T}$. 

\smallskip
\noindent
(c) From (\ref{TT'anosov1}) we infer: 
(i) if $us=0$ then $T'=T$, (ii) if $us>0$ then $T'>T$, (iii) if $us<0$ then $T'<T$.
{\hfill$\diamondsuit$}
\end{remark}

\subsection{Crossings}

For $p\in Y=\Gamma\backslash\H^2$ and $\xi,\zeta\in T^1_p(Y)\setminus\{0\}$ the angle
$\theta=\sphericalangle(\xi,\zeta)\in[0,\pi]$
between $\xi$ and $\zeta$ is given by
\[\cos\theta=\frac{g_{\Gamma,p}(\xi,\zeta)}{\|\xi\|_p\|\zeta\|_p}\]
for the natural Riemannian metric $g_\Gamma=(g_{\Gamma,p})_{p\in Y}$ on $Y$ and
associated norms $\|\cdot\|_p$ on $T_p(Y)$. Thus for $(p,\xi),(p,\zeta)\in\X=T^1(\Gamma\backslash \H^2)$ 
the angle $\theta=\sphericalangle(\xi,\zeta)\in[0,\pi]$ is obtained from
\[ \cos\theta=g_{\Gamma,p}(\xi,\zeta). \]

In the following lemma we consider a point $p$ on the surface $\Gamma\backslash\H^2$ 
and two unit tangent vectors $\xi$ and $\zeta$ at $p$. First we derive how the angle $\theta$ 
between $\xi$ and $\zeta$ yields a simple connection between the associated elements $g, h\in {\rm PSL}(2, \R)$: 
they are obtained from each other by applying a rotation $d_\theta$. 
In parts (b) and (c) we work out how this relation is affected if we pass 
to the reflected vectors $\xi'=-\xi$ and $\zeta'=-\zeta$. 

\begin{lemma}\label{gh-rela} (a) If $x=\Xi(p, \xi)$ and $y=\Xi(p, \zeta)$ 
for $(p, \xi), (p, \zeta)\in T^1(\Gamma\backslash\H^2)$, 
and if $\theta=\sphericalangle(\xi, \zeta)$, then $x=\Pi_\Gamma(g)$ and $y=\Pi_\Gamma(h)$ 
for some $g, h\in {\rm PSL}(2, \R)$ so that either $g=hd_\theta$ or $h=gd_\theta$. 
\smallskip 

\noindent 
(b) Furthermore, denote $\xi'=-\xi$ and $\zeta'=-\zeta$. 
If $x'=\Xi(p, \xi')$ and $y'=\Xi(p, \zeta')$, then $x'=\Pi_\Gamma(g')$ 
and $y'=\Pi_\Gamma(h')$ for $g'=gd_\pi$ and $h'=hd_\pi$. 
\smallskip 

\noindent 
(c) Either $g'=hd_{\theta+\pi}$ or $h'=gd_{\theta+\pi}$ holds. 
\end{lemma}

\noindent  
{\bf Proof\,:} (a) Fix $z \in\H^2$ such that $\pi_\Gamma(z )=p$. Next let
$\chi=\pi_\Gamma^{-1}:Y\supset V\rightarrow U\subset \H^2$ be a diffeomorphic 
coordinate chart satisfying $z \in U$ and $p\in {V} $. 
Such a set $V$ does exist since $\pi_\Gamma:\H^2\rightarrow \Gamma\backslash\H^2=Y$ is a local isometry.
Then $(d\chi^{-1})_z=(d\pi_\Gamma)_z:T_z(\H^2)\rightarrow T_p(Y)$ is an isometry.
Since $(p,\xi),(p,\zeta)\in T_p(Y)$, there are $\alpha,\beta\in T_z(\H^2)$ such that
\[ (d\pi_\Gamma)_z\alpha=\xi,\quad (d\pi_\Gamma)_z\beta=\zeta,\quad \|\alpha\|_z=\|\xi\|_p=1,\quad\mbox{and}\quad
\|\beta\|_z=\|\zeta\|_p=1,\]
where in general $\|\alpha\|_z=g_z(\alpha,\alpha)^{1/2}$ and $\|\xi\|_p=g_{\Gamma,p}(\xi,\xi)^{1/2}$.
Then the angle $\theta=\sphericalangle(\xi,\zeta)$ may be expressed as
\begin{equation}\label{cos}
   \cos(\theta)=g_{\Gamma,p}(\xi,\zeta)=g_{\Gamma,p}\Big((d\pi_\Gamma)_z\alpha,(d\pi_\Gamma)_z\beta\Big)
   =g_z(\alpha,\beta).
\end{equation} 
To use this, we first put $g=\Upsilon(z,\alpha)\in \PSL(2,\R)$ and  $h=\Upsilon(z,\beta)\in \PSL(2,\R)$
for the map $\Upsilon$ from (\ref{Upsidef}). Then the definition of $\Xi$ in the proof of theorem \ref{Xidef}
shows that 
\[ x=\Xi(p, \xi)=\Gamma g=\Pi_\Gamma(g)
   \quad\mbox{and}\quad y=\Xi(p, \eta)=\Gamma h=\Pi_\Gamma(h). \]  
Also by definition of $\Upsilon$, 
\begin{equation}\label{22} 
   {\cal D}g(i, i)=(z, \alpha)\quad\mbox{and}\quad {\cal D}h(i, i)=(z, \beta).
\end{equation} 
Next write $g=[G],\, h=[H]$ with 
$G=\Big(\scriptsize\begin{array}{cc} a_1&b_1\\c_1 & d_1
\end{array}\Big)\in\SL(2,\R),\,H=\Big(\begin{array}{cc}
a_2 & b_2 \\ c_2 & d_2\end{array}\Big)\in\SL(2,\R)$
and put $T=\Phi(g)\in\rm M\ddot ob(\H^2)$ as well as $S=\Phi(h)\in\rm M\ddot ob(\H^2)$. 
Owing to (\ref{calDexpl}) and (\ref{22}) we obtain the explicit relations
\[T(i)=\frac{a_1i+b_1}{c_1i+d_1}=z,\, T'(i)i=\frac{i}{(c_1i+d_1)^2}=\alpha,\,
S(i)=\frac{a_2i+b_2}{c_2i+d_2}=z,\, S'(i)i=\frac{i}{(c_2i+d_2)^2}=\beta.\]
It hence follows that 
\[{\rm Im\,} z=\frac{1}{|c_1i+d_1|^2}=\frac{1}{c_1^2+d_1^2}\quad \mbox{and}\quad
{\rm Im\,} z=\frac{1}{|c_2i+d_2|^2}=\frac{1}{c_2^2+d_2^2},\]
and thus
\begin{equation}\label{cd}
   c_1^2+d_1^2=c_2^2+d_2^2.
\end{equation}
Therefore $T(i)=S(i)$
leads to $(a_1i+b_1)(-c_1i+d_1)=(a_2i+b_2)(-c_2i+d_2)$,
so that
\begin{equation}\label{ac}
   a_1c_1+b_1d_1=a_2c_2+b_2d_2.
\end{equation}
Also
\begin{equation*}
   \alpha=\frac{2c_1d_1+i(d_1^2-c_1^2)}{(c_1^2+d_1^2)^2}
   \quad\mbox{and}\quad \beta=\frac{2c_2d_2+i(d_2^2-c_2^2)}{(c_2^2+d_2^2)^2}
\end{equation*}
by separation into real and imaginary parts. Then the preceding relations yield
\[a_1d_1^2=(1+b_1c_1)d_1=d_1+c_1(a_2c_2+b_2d_2-a_1c_1),\]
which means that
\begin{equation}\label{a1} 
   a_1=\frac{c_1(a_1c_1+b_1d_1)+d_1}{c_1^2+d_1^2}.
\end{equation}   
Similarly,
\[ b_1c_1^2=(a_1d_1-1)c_1=-c_1+d_1(a_2c_2=b_2d_2-b_1d_1) \]
leads to
\begin{equation}\label{b1}
 b_1=\frac{-c_1+(a_1c_1+b_1d_1)d_1}{c_1^2+d_1^2}.
\end{equation}
In the same way,
\[ a_2=\frac{d_2+c_2(a_2c_2+b_2d_2)}{c_2^2+d_2^2}\quad\mbox{and}\quad
   b_2=\frac{-c_2+d_2(a_2c_2+b-2d_2)}{c_2^2+d_2^2}. \] 
Now, back to (\ref{cos}), we have for $\theta=\sphericalangle (\xi,\zeta)$,
\[\cos(\theta)=g_z(\alpha,\beta)=\frac{1}{({\rm Im\,} z)^2}\,(\alpha_1\beta_1+\alpha_2\beta_2)
=\frac{1}{(c_1^2+d_1^2)^2}\,(4c_1c_2d_1d_2+(d_1^2-c_1^2)(d_2^2-c_2^2))\]
in terms of the matrix coefficients. Using (\ref{cd}) we obtain
\begin{eqnarray}\notag \label{cosi-rel} 
   \cos^2\Big(\frac{\theta}2\Big)=\frac12(1+\cos(\theta))
   & = & \frac12\Big(\frac{4c_1c_2d_1d_2+(d_1^2-c_1^2)(d_2^2-c_2^2)+(c_1^2+d_1^2)^2}{(c_1^2+d_1^2)^2}\Big)\\
   & = & \frac{(c_1c_2+d_1d_2)^2}{(c_1^2+d_1^2)^2}.
\end{eqnarray}
Hence as a consequence of $(c_1^2+d_1^2)^2=(c_1^2+d_1^2)(c_2^2+d_2^2)$ we obtain 
\begin{equation}\label{sini-rel} 
   \sin^2\Big(\frac{\theta}{2}\Big)=1-\cos^2\Big(\frac{\theta}{2}\Big)
   =\frac{(c_1^2+d_1^2)(c_2^2+d_2^2)-(c_1 c_2+d_1 d_2)^2}{(c_1^2+d_1^2)^2}
   =\frac{(d_1 c_2-c_1 d_2)^2}{(c_1^2+d_1^2)^2}.
\end{equation} 
Now we need to distinguish several cases related to the sign of the square root 
in (\ref{cosi-rel}) and (\ref{sini-rel}). 
\smallskip

\noindent
\underline{Case 1:} $\cos(\frac{\theta}{2})=\frac{c_1 c_2+d_1 d_2}{c_1^2+d_1^2}$ 
and $\sin(\frac{\theta}{2})=\frac{d_1 c_2-c_1 d_2}{c_1^2+d_1^2}$. Then $G=HD_\theta$, since
\begin{eqnarray*} 
   HD_\theta  =  \Bigg(\begin{array}{cc} a_2\,\frac{c_1 c_2+d_1 d_2}{c_1^2+d_1^2}
   -b_2\,\frac{d_1 c_2-c_1 d_2}{c_1^2+d_1^2}
   & a_2\,\frac{d_1 c_2-c_1 d_2}{c_1^2+d_1^2}+b_2\,\frac{c_1 c_2+d_1 d_2}{c_1^2+d_1^2} 
   \\[1ex] c_2\,\frac{c_1 c_2+d_1 d_2}{c_1^2+d_1^2}-d_2\,\frac{d_1 c_2-c_1 d_2}{c_1^2+d_1^2}
   & c_2\,\frac{d_1 c_2-c_1 d_2}{c_1^2+d_1^2}+d_2\,\frac{c_1 c_2+d_1 d_2}{c_1^2+d_1^2}\end{array}\Bigg),
\end{eqnarray*} 
and (\ref{ac}) together with (\ref{a1}) yield 
\begin{eqnarray*} 
   a_2\,\frac{c_1 c_2+d_1 d_2}{c_1^2+d_1^2}-b_2\,\frac{d_1 c_2-c_1 d_2}{c_1^2+d_1^2}
   & = & \frac{c_1(a_2 c_2+b_2 d_2)+d_1(a_2 d_2-b_2 c_2)}{c_1^2+d_1^2}
   \\ & = & \frac{c_1(a_1 c_1+b_1 d_1)+d_1}{c_1^2+d_1^2}=a_1, 
\end{eqnarray*} 
whereas (\ref{ac}) and (\ref{b1}) show that 
\begin{eqnarray*} 
   a_2\,\frac{d_1 c_2-c_1 d_2}{c_1^2+d_1^2}+b_2\,\frac{c_1 c_2+d_1 d_2}{c_1^2+d_1^2}
   & = & \frac{c_1(b_2 c_2-a_2 d_2)+d_1(a_2 c_2+b_2 d_2)}{c_1^2+d_1^2}
   \\ & = & \frac{-c_1+d_1(a_1 c_1+b_1 d_1)}{c_1^2+d_1^2}=b_1.
\end{eqnarray*} 
Similarly, due to $c_1^2+d_1^2=c_2^2+d_2^2$, 
\[ c_2\,\frac{c_1 c_2+d_1 d_2}{c_1^2+d_1^2}-d_2\,\frac{d_1 c_2-c_1 d_2}{c_1^2+d_1^2}
   =\frac{c_1(c_2^2+d_2^2)}{c_1^2+d_1^2}=c_1, \] 
and furthermore 
\[ c_2\,\frac{d_1 c_2-c_1 d_2}{c_1^2+d_1^2}+d_2\,\frac{c_1 c_2+d_1 d_2}{c_1^2+d_1^2}
   =\frac{d_1(c_2^2+d_2^2)}{c_1^2+d_1^2}=d_1. \]
Thus indeed $G=HD_\theta$, and consequently $g=hd_\theta$. 
\smallskip

\noindent
\underline{Case 2:} $\cos(\frac{\theta}{2})=\frac{c_1 c_2+d_1 d_2}{c_1^2+d_1^2}$ 
and $\sin(\frac{\theta}{2})=-\frac{d_1 c_2-c_1 d_2}{c_1^2+d_1^2}$. Similarly, we obtained  $G=HD_{-\theta}$
after a shot calculation. Hence $g=hd_{-\theta}$ and thus $h=gd_\theta$. 
\smallskip

\noindent
\underline{Case 3:} $\cos(\frac{\theta}{2})=-\frac{c_1 c_2+d_1 d_2}{c_1^2+d_1^2}$  
and $\sin(\frac{\theta}{2})=\frac{d_1 c_2-c_1 d_2}{c_1^2+d_1^2}$. Then $G=-HD_{-\theta}$, 
hence $g=hd_{-\theta}$ and thus $h=gd_\theta$. 
\smallskip

\noindent
\underline{Case 4:} $\cos(\frac{\theta}{2})=-\frac{c_1 c_2+d_1 d_2}{c_1^2+d_1^2}$ 
and $\sin(\frac{\theta}{2})=-\frac{d_1 c_2-c_1 d_2}{c_1^2+d_1^2}$. Then $G=-HD_\theta$, 
hence $g=hd_\theta$, completing the proof of (a). 

\smallskip
\noindent
(b) We continue to use the notation from (a). 
Defining $\alpha'=-\alpha\in T_z(\H^2)$ and $\beta'=-\beta\in T_z(\H^2)$, 
we have ${(d\pi_\Gamma)}_z\alpha'=-{(d\pi_\Gamma)}_z\alpha=-\xi=\xi'$ 
and ${(d\pi_\Gamma)}_z\beta'=\eta'$ for unit vectors $\alpha'$ and $\beta'$, 
by the linearity of the tangent map ${(d\pi_\Gamma)}_z: T_z(\H^2)\to T_p(\Gamma\backslash\H^2)$. 
Then by definition of $\Xi$ in the proof of theorem \ref{Xidef} 
we obtain $x'=\Xi(p, \xi')=\Pi_\Gamma(\hat{g})$ and $y'=\Xi(p, \eta')=\Pi_\Gamma(\hat{h})$ 
for $\hat{g}=\Upsilon(z, \alpha')$ and $\hat{h}=\Upsilon(z, \beta')$. 
But lemma \ref{LgDg} shows that $\hat{g}=\Upsilon(z, \alpha')=\Upsilon(z, -\alpha)
=\Upsilon(z, \alpha)j=gd_\pi=g'$, and similarly $\hat{h}=h'$, 
due to $j=[J]$ for $J=\scriptsize \Big(\begin{array}{cc} 0 & 1 \\ -1 & 0\end{array}\Big)=D_\pi$. 

\smallskip
\noindent
(c) By (b) and (a) we have $g'=gd_\pi=hd_\theta d_\pi=h d_{\theta+\pi}$ 
or $h'=hd_\pi=gd_\theta d_\pi=gd_{\theta+\pi}$. 
{\hfill$\Box$}\bigskip

\begin{remark}\rm
In the setting of lemma \ref{gh-rela}(a), either $\Gamma h=\Gamma g d_\theta$ or $\Gamma g=\Gamma h d_\theta$
holds for any $g,h\in\PSL(2,\R)$ such that $x=\Gamma g, y=\Gamma h$.  {\hfill$\diamondsuit$}
\end{remark}

The next result is a converse statement to lemma \ref{gh-rela}(a).

\begin{lemma}\label{inverlm} Let $x=\Xi(p,\xi)$ and
$y=\Xi(q,\zeta)$ for
$(p,\xi),(q,\zeta)\in T^1(\Gamma\backslash \H^2)$.
If $x=\Pi_\Gamma(g), y=\Pi_\Gamma(h)$, and
$g=hd_\theta$ for $g,h\in\PSL(2,\R), \theta\in ]-\pi,\pi[$, then
$p=q$. In addition, if all elements in
 $\Gamma\setminus\{e\}$ are hyperbolic, then $\sphericalangle (\xi,\zeta)=|\theta|$.
\end{lemma}
\noindent
{\bf Proof\,:} Write 
$p=\pi_\Gamma(z), q=\pi_\Gamma(z')$ for $z,z'\in \H^2$ and $g=\Upsilon(z,\alpha), h=\Upsilon(z',\beta)$
for $(z,\alpha), (z',\beta)\in T^1\H^2$. We
are going to show that $z=z'$.
Writing $g=[G], h=[H]$ for $G=\Big(\scriptsize\begin{array}{cc} a & b \\ c & d
\end{array}\Big)\in\SL(2,\R), H=\Big(\begin{array}{cc} a' & b' \\ c' & d'
\end{array}\Big)\in\SL(2,\R)$, by the definition of $\Upsilon$ from (\ref{Upsidef})
we have $ {\cal D}g(i,i)=(z,\alpha)$ and ${\cal D}h (i,i)=(z',\beta)$. 
Using the definition of ${\cal D}$ from (\ref{calDexpl}), it follows that
\begin{eqnarray}
  z & = &\Phi(g)(i)=\frac{ai+b}{ci+d}=\frac{ac+bd}{c^2+d^2}+i\frac{1}{c^2+d^2},
  \label{z}
  \\ 
  z' & = & \Phi(h)(i)=\frac{a'i+b'}{c'i+d'}=\frac{a'c'+b'd'}{c'^2+d'^2}+i\frac{1}{c'^2+d'^2}.
  \label{z'}
\end{eqnarray}
By the assumption we have $g=h d_\theta$, and without loss of generality 
we can assume that $G=HD_\theta$. A short calculation shows that
\begin{eqnarray*}
  a & = & a'\cos\Big(\frac{\theta}2\Big)-b'\sin\Big(\frac{\theta}2\Big),\quad 
  b\ =\ a'\sin\Big(\frac{\theta}2\Big)+b'\cos\Big(\frac{\theta}2\Big), 
  \\ c & = & c'\cos\Big(\frac{\theta}2\Big)-d'\sin\Big(\frac{\theta}2\Big),\quad 
  d\ =\ c'\sin\Big(\frac{\theta}2\Big)+d'\cos\Big(\frac{\theta}2\Big).
\end{eqnarray*}
This implies that
\begin{eqnarray}
  c^2+d^2= c'^2+d'^2\quad \mbox{and}\quad
  ac+bd=a'c'+b'd',
\end{eqnarray}
hence $z=z'$ due to (\ref{z}) and (\ref{z'}), and thus $p=q$.
It remains to show that  $\sphericalangle(\xi,\zeta)=|\theta|$ if every element in $\Gamma\backslash\{e\}$ is hyperbolic. 
To see this, let $\sphericalangle(\xi,\zeta)=\phi \in [0,\pi]$. 
By lemma \ref{gh-rela}(a), either $\Gamma g=\Gamma h d_\phi$ or $\Gamma g=\Gamma h d_{-\phi}$ holds. 
\underline{Case 1:} $\Gamma g=\Gamma h d_\phi$. 
Then $\Gamma h d_\phi=\Gamma h d_\theta$ yields 
$h^{-1}\gamma h=d_{\phi-\theta}$ for some $\gamma\in\Gamma$ and hence $\tr(\gamma)
=\tr(h^{-1}\gamma h)=\tr(d_{\phi-\theta})=2\cos(\frac{\phi-\theta}{2}) \leq 2$.
By assumption, we have $\gamma=e$ and thus $\varphi=\theta$.
\underline{Case 2:} $\Gamma h=\Gamma k d_{-\phi}$. Here similarly we obtain
 $\phi=-\theta$. In summary, we have shown that $\phi=|\theta|$. 
{\hfill$\Box$}\bigskip

Next we give a necessary and sufficient condition for 
crossings. 

\begin{theorem}[Crossings] 
Suppose that all elements of $\Gamma\setminus\{e\}$ 
are hyperbolic and let  ${\mathtt x_1}=(p_1,\xi_1),{\mathtt x_2}=(p_2,\xi_2)\in \X=T^1(\Gamma\backslash\H^2)$ be given. 
The orbits of ${\mathtt x_1}$ and ${\mathtt x_2}$  under the geodesic flow $(\varphi_t^\X)_{t\in\R}$ 
have an intersection in configuration space at an angle $\theta\in ]0,\pi[$ if and only if
there are $t_1,t_2\in\R$ such that 
\begin{equation}\label{g1g2}
   \mbox{either}\quad \Gamma g_1a_{t_1}=\Gamma g_2 a_{t_2}d_\theta   
   \quad \mbox{or}\quad \Gamma g_2a_{t_2}=\Gamma g_1 a_{t_1}d_\theta
\end{equation}
holds,
where $g_1,g_2\in\PSL(2,\R)$ are such that
$\Xi(\mathtt x_1)=\Gamma g_1, \Xi(\mathtt x_2)=\Gamma g_2$.
\end{theorem}
\noindent
{\bf Proof\,:} Suppose that the two orbits of ${\mathtt x_1}$ 
and ${\mathtt x_2}$ have an intersection in configuration space at a point $p\in\Gamma\backslash\H^2$
and at the angle $\theta\in ]0,\pi[$. Then there exist $t_1,t_2\in\R$ such that
$\varphi_{t_1}^\X({\mathtt x_1})=(p,\xi),\varphi^\X_{t_2}({\mathtt x_2})=(p,\zeta)$, and 
$\theta=\sphericalangle(\xi,\zeta)$. 
Noting that $\Gamma g_1a_{t_1}=\varphi_{t_1}^X(\Gamma g_1)=\Xi(p,\xi)$ and $\Gamma g_2a_{t_2}
=\varphi_{t_2}^X(\Gamma g_2)=\Xi(p,\zeta)$,
we see that (\ref{g1g2}) holds by lemma \ref{gh-rela}(a).   
Conversely, suppose that (\ref{g1g2}) holds for some $t_1,t_2\in\R$. 
Letting  $\varphi_{t_1}^\X({\mathtt x_1})=(p,\xi)$ and
$\varphi_{t_2}^\X({\mathtt x_2})=(q,\zeta)$, we have 
correspondingly $\varphi^X_{t_1}(\Gamma g_1)=\Gamma g_1a_{t_1}= \Xi(p,\xi)$ as well as
$\varphi^X_{t_2}(\Gamma g_2)=\Gamma g_2 a_{t_2}=\Xi(q,\zeta)$. 
Now we apply lemma \ref{inverlm} to obtain $p=q$ and 
$\sphericalangle(\xi,\zeta)=\theta$, i.e., the orbits of ${\mathtt x_1}$
and ${\mathtt x_2}$ intersect in configuration space at the point $p\in \Gamma\backslash\H^2$ and the angle $\theta$.
{\hfill$\Box$}\bigskip

\begin{figure}[ht]
\begin{center}
\begin{minipage}{0.5\linewidth}
   \centering
   \includegraphics[angle=0,width=0.7\linewidth]{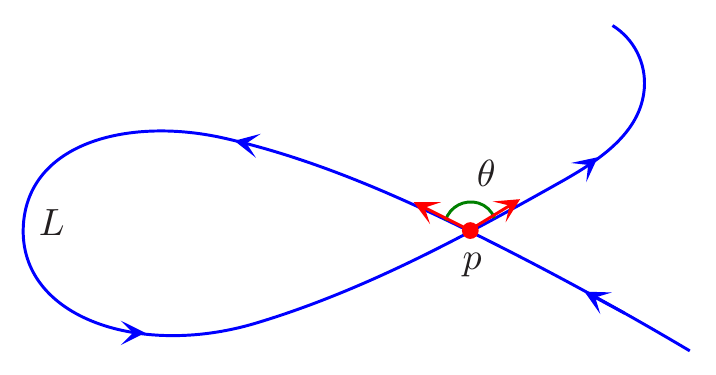}
\end{minipage}
\end{center}
\caption{An orbit with a self-crossing in configuration space}\label{self_crossing}
\end{figure}

Figure \ref{self_crossing} illustrates the next result. 

\begin{theorem}[Self-crossings]\label{selfcthm} 
Suppose that all elements of $\Gamma\setminus\{e\}$ are hyperbolic 
and let  $\tau\in\R,\, L>0,\,\theta\in\,]0,\pi[$, and ${\mathtt x}=(p,\xi)\in \X$ be given. 
The orbit of ${\mathtt x}$ under the geodesic flow 
${(\varphi^\X_t)}_{t\in\R}$ crosses itself in configuration space at the 
time $\tau$, at the angle $\theta$, and creates a loop of length $L$ if and only if
\[ \mbox{either}\quad \Gamma ga_{\tau+ L}=\Gamma g a_\tau d_\theta \quad\mbox{or}\quad
   \Gamma g_{\tau+L}=\Gamma g a_{\tau} d_{-\theta} \] 
holds for any $g\in\PSL(2,\R), \Gamma g=\Xi({\mathtt x})$.
\end{theorem}
\noindent
{\bf Proof\,:} Fix $g\in\PSL(2,\R)$ such that $\Gamma g=\Pi_\Gamma(g)=\Xi({\mathtt x})$. 
We apply the preceding theorem for ${\mathtt x_1}={\mathtt x_2}={\mathtt x}$
and $t_1=\tau, t_2=\tau+L$ to obtain a self-crossing in configuration space at the time $\tau$.
The self-crossing creates a loop which starts at the time $\tau$
and ends at the time $\tau+L$. So the length of the loop is $L$; see
Figure \ref{self_crossing}.
{\hfill$\Box$}\bigskip

The relation between the loop length and the self-crossing angle is illustrated by the following result 
which is a necessary condition for self-crossings; formula (\ref{nevcam}) below 
had already been derived in \cite[p.~133]{Sieber1}. 

\begin{corollary}\label{periodandangle}
Suppose that all elements of $\Gamma\setminus\{e\}$ are hyperbolic. 
If an orbit of the geodesic flow crosses itself in configuration space at 
an angle $\theta$ and creates a loop of length $L$, then
\begin{itemize}
\item[(a)] there is $\rho>0$ such that 
\begin{equation}\label{nevcam}
   \cosh\Big(\frac{\rho}{2}\Big)=\cosh\Big(\frac{L}{2}\Big)\cos\Big(\frac{\theta}{2}\Big);
\end{equation} 
\item[(b)] 
\begin{equation}
   e^{-L}<\cos^2\Big(\frac{\theta}{2}\Big). 
\end{equation}
\end{itemize}
\end{corollary}
\noindent
{\bf Proof\,:} Let the orbit of ${\mathtt x}\in\X$ satisfy the assumption and put 
$\Gamma g=\Xi({\mathtt x})$. (a) According to theorem \ref{selfcthm} for $\tau=0$, 
\[\mbox{either}\quad \Gamma ga_L=\Gamma g d_{\theta} \quad \mbox{or}\quad \Gamma g a_L=\Gamma gd_{-\theta}\quad \mbox{holds}.\]
\underline{Case 1:} $\Gamma g a_L=\Gamma gd_\theta$. 
Writing $ga_{L}=\gamma gd_\theta$ for some $\gamma\in\Gamma \setminus\{e\}$
yields $\gamma=ga_{L}d_{-\theta}g^{-1}$, and accordingly 
$\tr(\gamma)=\tr(a_Ld_{-\theta})$.  
Recall lemma \ref{exgeodper} and let $\rho>0$ be such that  $e^{\rho/2}+e^{-\rho/2}={\rm tr}(\gamma)$. Then 
$\cosh(\frac\rho2)=\frac{1}{2}{\rm tr}(a_{L} d_{-\theta})$. On the other hand, 
\[ A_{L} D_{-\theta}=\Bigg(\begin{array}{cc} e^{L/2}\cos(\frac{\theta}{2}) & -e^{L/2}\sin(\frac{\theta}{2}) 
   \\[1ex] e^{-L/2}\sin(\frac{\theta}{2}) & e^{-L/2}\cos(\frac{\theta}{2})\end{array}\Bigg) \] 
leads to ${\rm tr}(a_{L} d_{-\theta})=|{\rm tr}(A_{L}D_{-\theta})|=(e^{L/2}+e^{-L/2})\cos(\frac{\theta}{2})$ 
by definition of the trace. \underline{Case 2:} $\Gamma ga_L=\Gamma gd_{-\theta}$. Similarly we write $ga_L=\gamma gd_{-\theta}$ for $\gamma\in\Gamma\setminus\{e\}$. Then $\gamma=ga_Ld_{\theta}g^{-1}$ yields 
${\rm tr}(\gamma)=e^{\rho/2}+e^{-\rho/2}={\rm tr}(a_{L}d_\theta )$ for some $\rho>0$. Since 
\[ A_L D_{\theta} =\Bigg(\begin{array}{cc} e^{L/2}\cos(\frac{\theta}{2}) & e^{L/2}\sin(\frac{\theta}{2}) 
   \\[1ex] -e^{-L/2}\sin(\frac{\theta}{2}) & e^{-L/2}\cos(\frac{\theta}{2})\end{array}\Bigg), \] 
we obtain ${\rm tr}(a_{L}d_\theta )=(e^{L/2}+e^{-L/2})\cos(\frac{\theta}{2})$, 
completing the proof of (\ref{nevcam}). (b) In particular,
noting that $(e^{L/2}+e^{-L/2})|\cos(\frac{\theta}{2})|=e^{\rho/2}+e^{-\rho/2}>2$, 
we have \[\Big|\cos\Big(\frac{\theta}{2}\Big)\Big|>\frac{2}{e^{L/2}+e^{-L/2}}>e^{-L/2}\]
which
yields (\ref{nevcam}). 
\smallskip
{\hfill$\Box$}\bigskip

The next observation allows us to find a self-crossing orbit with a prescribed crossing angle. 

\begin{lemma} 
Suppose that all elements of $\Gamma\setminus\{e\}$ are hyperbolic. 
For every $\gamma\in\Gamma\setminus\{e\}$ and $\theta \in\,]0,\pi[$
there is an orbit of the geodesic flow that crosses itself in configuration space at
the angle $\theta$ and creates a loop whose length $l_\phi(\gamma)$ is determined by
\begin{eqnarray}\label{lLeq}
   \cosh\Big(\frac{l(\gamma)}2\Big)=\cosh\Big(\frac{l_\phi(\gamma)}2\Big)\cos\Big(\frac{\theta}{2}\Big);
\end{eqnarray}
here $l(\gamma)>0$ is the number which is determined by 
\[ e^{l(\gamma)/2}+e^{-l(\gamma)/2}=\tr(\gamma). \]
\end{lemma}
\noindent 
{\bf Proof\,:} Since $\frac{\cosh(\frac{l(\gamma)}{2})}{\cos(\frac{\theta}{2})}>1$, 
a unique solution $l_\phi(\gamma)>0$ to (\ref{lLeq}) does exist. Denoting $L=l_\phi(\gamma)$, 
we shall find a $g\in\PSL(2,\R)$ such that 
\begin{equation}\label{gtheta}
   \gamma ga_L=gd_\theta.
\end{equation}
First we apply lemma \ref{exgeodper} to write $\gamma =h^{-1}a_{l(\gamma)}h$
for some $h\in\PSL(2,\R)$ and rewrite (\ref{gtheta}) as
$a_{l(\gamma)}k=k d_\theta a_{-L}$ for $k=hg$.
It suffices to find 
$K=\scriptsize\Big(\begin{array}{cc} a&b\\ c&d \end{array}\Big)\in\SL(2,\R)$
satisfying the equation $A_{l(\gamma)}K=KD_\theta A_{-L}$, which means that 
\begin{eqnarray}
   \label{eq1} 
   a(e^{-L/2}\cos(\tfrac{\theta}{2})-e^{l(\gamma)/2})-be^{-L/2}\sin(\tfrac{\theta}{2})&=&0, 
   \\
   \label{eq2}
   a e^{L/2} \sin(\tfrac{\theta}{2})+b(e^{L/2} \cos(\tfrac{\theta}{2})-e^{l(\gamma)/2})&=&0,
   \\ 
   \label{eq3}
   c(e^{-L/2}\cos(\tfrac{\theta}{2})-e^{-l(\gamma)/2})-d e^{-L/2} \sin(\tfrac{\theta}{2})&=&0,
   \\ 
   \label{eq4}
   ce^{L/2}\sin(\tfrac{\theta}{2}) +d (e^{L/2}\cos(\tfrac{\theta}{2})-e^{-l(\gamma)/2}) &=&0.
\end{eqnarray} 
Note that the equations in (\ref{eq1})\&(\ref{eq2}) are equivalent and so are 
the equations in (\ref{eq3})\&(\ref{eq4}). Furthermore, it is obvious that
$e^{-L/2}\cos(\tfrac{\theta}{2})-e^{l(\gamma)/2}\ne 0$. 
Therefore if we take any $b,c\in\R$ and let
\[ a=\frac{be^{-L/2}\sin(\tfrac{\theta}{2})}
   {e^{-L/2}\cos(\tfrac{\theta}{2})-e^{l(\gamma)/2}},
   \quad d=\frac{c(e^{-L/2}\cos(\tfrac{\theta}{2})-e^{-l(\gamma)/2})}
   {e^{-L/2} \sin(\tfrac{\theta}{2})}
\]
then all the four equations in (\ref{eq1})-(\ref{eq4})
are solved. Now the special choice
\[ b=e^{-L/2}\cos(\tfrac{\theta}{2})-e^{l(\gamma)/2}, \quad 
   c=\frac{1}{e^{l(\gamma)/2}-e^{-l(\gamma)/2}} \]
yields $\det(K)=ad-bc=1$ after a short calculation.
Hence $g=h^{-1}k$ for $k=[K]$ is a solution to
the equation (\ref{gtheta}) and by theorem \ref{selfcthm},
the orbit through $x=\Pi_\Gamma(g)$ is as desired. 
{\hfill$\Box$}\bigskip 

\subsection{Existence of a partner orbit and action difference of a Sieber-Richter pair}

\begin{figure}[ht]
\begin{center}
\begin{minipage}{0.8\linewidth}
   \centering
   \includegraphics[angle=0,width=0.7\linewidth]{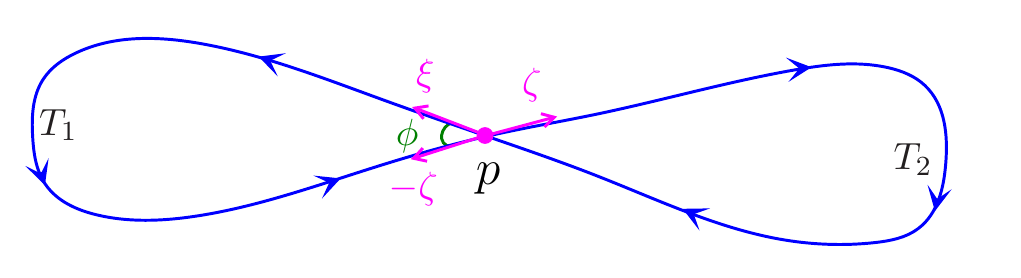}
\end{minipage}
\end{center}
\caption{Small-angle self-crossing in configuration space}\label{smallangle}
\end{figure}

\begin{theorem}[Existence of a partner orbit I]\label{existthm}
If a periodic orbit of the geodesic flow ${(\varphi^\X_t)}_{t\in\R}$ on $\X=T^1(\Gamma\backslash\H^2)$
with the period $T\geq 1$ crosses itself in configuration space at a time $T_1\in\,]0,T[$
and at an angle $\theta$ such that $0<\phi<\frac13$
for $\phi=\pi-\theta$, 
then there is another periodic orbit of the geodesic flow (called a partner orbit)
which remains $9|\sin(\phi/2)|$-close to the original one.
Furthermore, $T'<T$ for the period of the partner orbit and
\begin{eqnarray}\label{actdiff}
   \Big|\frac{T'-T}{2}-\ln\Big(1-(1+e^{-T_1})(1+e^{-(T-T_1)})\sin^2(\phi/2)\Big)\Big|\leq 12\sin^2(\phi/2)e^{-T}.
\end{eqnarray}
The original orbit and its partner are called a Sieber-Richter pair. 
\end{theorem}
\noindent
{\bf Proof\,:} Let the orbit of $(p, \xi)\in {\cal X}=T^1(\Gamma\backslash\H^2)$ 
be $T$-periodic ($T$ is the prime period)
and such that it has a self-crossing of angle $\theta$ 
in configuration space at the time $T_1\in \,]0, T[$, i.e., we have 
\begin{equation}\label{seya}
   \varphi_{T_1}^{{\cal X}}(p, \xi)=(p, \zeta),
   \quad\varphi_{T_2}^{{\cal X}}(p, \zeta)=(p, \xi),\quad\mbox{and}\quad\sphericalangle (\xi,\zeta)=\theta,
\end{equation} 
where $T=T_1+T_2$; see Figure \ref{smallangle}.

\noindent
In addition, assume that $|\phi|<\frac13$ with $\phi=\pi-\theta$. Then in particular 
\begin{equation}\label{sinphi/2} 
   \Big|\sin\Big(\frac{\phi}{2}\Big)\Big|\le\frac{|\phi|}{2}
   <\frac16
\end{equation} 
holds.
Denote $x=\Xi(p, \xi)\in X=\Gamma\backslash {\rm PSL}(2, \R)$ 
and $y=\Xi(p, \zeta)\in X$. Then, according to lemma \ref{gh-rela}(a), 
we may write $x=\Pi_\Gamma(g)$ and $y=\Pi_\Gamma(h)$ 
for some $g, h\in {\rm PSL}(2, \R)$ so that either $g=hd_\theta$ or $h=gd_\theta$.
Due to (\ref{geodflcalX}) and (\ref{seya}) we obtain 
\begin{equation}\label{fste1n} 
   \varphi_{T_1}^X(x)=(\Xi\circ\varphi_{T_1}^{{\cal X}}\circ\Xi^{-1})(x)
   =(\Xi\circ\varphi_{T_1}^{{\cal X}})(p, \xi)=\Xi(p, \zeta)=y.
\end{equation} 
Furthermore, 
\begin{equation}\label{fste2n} 
   \varphi_{T_2}^X(y)=(\Xi\circ\varphi_{T_2}^{{\cal X}}\circ\Xi^{-1})(y)
   =(\Xi\circ\varphi_{T_2}^{{\cal X}})(p, \zeta)=\Xi(p, \xi)=x,
\end{equation} 
and hence in particular $\varphi_T^X(x)=\varphi_{T_2}^X(\varphi_{T_1}^X(x))=\varphi_{T_2}^X(y)=x$. 
By (\ref{fste1n}), (\ref{fste2n}), and the definition of $(\varphi_t^X)$  we have 
\[ \Pi_\Gamma(g a_{T_1})=\varphi_{T_1}^X(x)=y=\Pi_\Gamma(h)
   \quad\mbox{and}\quad\Pi_\Gamma(h a_{T_2})=\varphi_{T_2}^X(y)=x=\Pi_\Gamma(g). \] 
Hence there are $\gamma_1, \gamma_2\in\Gamma$ such that 
\begin{equation}\label{flukn} 
   ga_{T_1}=\gamma_1 h\quad\mbox{and}\quad ha_{T_2}=\gamma_2 g.
\end{equation} 
Next let $\xi'=-\xi$ and $\zeta'=-\zeta$. 
If $x'=\Xi(p, \xi')$ and $y'=\Xi(p, \zeta')$, then $x'=\Pi_\Gamma(g')$ 
and $y'=\Pi_\Gamma(h')$ for $g'=gd_\pi$ and $h'=hd_\pi$ by lemma \ref{gh-rela}(b). 
In addition, either $g'=hd_{\theta+\pi}$ or $h'=gd_{\theta+\pi}$ 
by lemma \ref{gh-rela}(c). Henceforth, we are going to assume that $g'=hd_{\theta+\pi}$, 
since the case where $h'=gd_{\theta+\pi}$ can be treated analogously. 
In view of (\ref{fste1n}) and (\ref{fste2n}) we may apply lemma \ref{revpropX} 
(see the proof of lemma \ref{gh-rela}(a)\&(b)) to deduce that 
\begin{equation}\label{fste3n} 
   \varphi_{T_1}^X(y')=x'\quad\mbox{and}\quad
   \varphi_{T_2}^X(x')=y'. 
\end{equation} 
As above it follows that there are $\gamma_1', \gamma_2'\in\Gamma$ such that 
\begin{equation}\label{flukn2}
   h'a_{T_1}=\gamma_1'g'\quad\mbox{and}\quad g'a_{T_2}=\gamma_2'h'.
\end{equation}The angle complementary to $\theta=\sphericalangle(\xi, \eta)$ is denoted
\[ \phi=\sphericalangle(\xi, \eta')=\sphericalangle(\xi', \eta)=\pi-\theta, \] 
cf.~Figure \ref{smallangle}, so that $g'=hd_{2\pi-\phi}$. 
Since $D_{2\pi-\phi}=-D_{-\phi}$ we have $d_{2\pi-\phi}=d_{-\phi}$ 
and hence $g'=hd_{-\phi}$. From $g'=gd_\pi$ and $h'=hd_\pi$ we obtain 
\begin{equation}\label{g=h'} 
   gd_\phi=gd_{2\pi+\phi}=g'd_{\pi+\phi}=hd_{-\phi}d_{\pi+\phi}=hd_\pi=h'.
\end{equation} 
By (\ref{sinphi/2}), we have 
\begin{equation}\label{cosphi/2}
   \cos\Big(\frac{\phi}{2}\Big)>\frac{5}{6}.
\end{equation} 
Thus using lemma \ref{decompo}(c) we can write 
\begin{equation}\label{dexpress} 
   d_{\phi}=b_sc_u a_\tau,
\end{equation} 
where 
\[ \tau=-2\ln(\cos(\phi/2)),\quad s=\tan(\phi/2),\quad u=-\sin(\phi/2)\cos(\phi/2). \]  
Then 
\begin{equation}\label{tausu-bd}  
   |s|=|\tan(\phi/2)|\le \frac32|\sin(\phi/2)|=:\eps,
\ |u|=|\sin(\phi/2)\cos(\phi/2)|\le |\sin(\phi/2)|<\eps,
\end{equation} 
and 
\[ |\tau|=|\ln(1-\sin^2(\phi/2))|\le 2\sin^2(\phi/2)
   \le \frac12\, \eps^2, \]
due to $|\ln(1+z)|\le 2|z|$ for $|z|\le 1/2$. Denote 
\[ \hat{x}=\Gamma g b_{s}\in X\quad\mbox{and}\quad\tilde{y}=\Gamma h' a_{-\tau}\in X. \] 
Then $\hat{x}\in W^s_\eps(x)$; see definition \ref{locsu-mfk}. By (\ref{g=h'}) and (\ref{dexpress}),
\[  \hat{x}=\Gamma g b_{s}=\Gamma h' d_{-\phi} b_{s}=\Gamma h' a_{-\tau} c_{-u}, \] 
whence we also obtain $\hat{x}\in W^u_\eps(\tilde{y})$. Therefore we can employ 
the shadowing lemma (theorem \ref{shadlemII}) to deduce that 
\begin{equation}\label{shadcons1} 
   d_X(\varphi^X_t(x), \varphi^X_t(\hat{x}))<\eps e^{-t}
   \quad\mbox{for all}\quad t\in [0, \infty[
\end{equation} 
and 
\begin{equation}\label{shadcons2} 
   d_X(\varphi^X_t(\tilde{y}), \varphi^X_t(\hat{x}))<\eps e^t 
   \quad\mbox{for all}\quad t\in\, ]-\infty, 0].
\end{equation} 
Then by (\ref{shadcons2}) and corollary \ref{distancePSL}(a) for $t\in \,]-\infty, 0]$, 
\begin{eqnarray}\label{versush} 
   d_X(\varphi^X_t(y'), \varphi^X_t(\hat{x}))
   & \le & d_X(\varphi^X_t(y'), \varphi^X_t(\tilde{y}))+d_X(\varphi^X_t(\tilde{y}), \varphi^X_t(\hat{x}))
   \nonumber
   \\ & \le & \inf_{\gamma\in\Gamma} d_{\G}(h' a_t, \gamma h' a_{-\tau} a_t)+\eps e^t
   \nonumber
   \\ & \le & d_{\G}(a_\tau, e)+\eps=\frac{1}{\sqrt{2}}\,|\tau|+\eps
    < 2\eps. 
\end{eqnarray} 
For 
\begin{equation}\label{goverACL} 
   \hat{y}=\varphi^X_{-T_2}(\hat{x})=\Gamma gb_{s} a_{-T_2}\in X
\end{equation} 
we are going to verify the assumptions of the Anosov closing lemma
(see theorem \ref{anosov1}). To begin with, 
\[ \varphi^X_T(\hat{y})=\varphi^X_{T-T_2}(\hat{x})=\varphi^X_{T_1}(\hat{x})=\Gamma gb_{s} a_{T_1}. \]
Then (\ref{flukn}), $g'=hd_{-\phi}$, and (\ref{flukn2}) yield
\begin{eqnarray*} 
   gb_{s} a_{T_1} & = & \gamma_1 h a_{-T_1} b_{s} a_{T_1}
   =\gamma_1 g' d_\phi b_{se^{-T_1}}=\gamma_1 \gamma'_2 h' a_{-T_2} d_\phi b_{se^{-T_1}}
   =\gamma_1 \gamma'_2 g d_\phi a_{-T_2} d_\phi b_{se^{-T_1}}
   \\ & = & \gamma_1 \gamma'_2 g b_{s} a_{-T_2} a_{T_2} b_{-s} d_\phi a_{-T_2} d_\phi b_{se^{-T_1}}, 
\end{eqnarray*} 
using $b_sa_{T_1}=a_{T_1}b_{se^{-T_1}}$,
so that 
\[ \varphi^X_T(\hat{y})=\Gamma g b_{s} a_{-T_2} (a_{T_2} b_{-s} d_\phi a_{-T_2} d_\phi b_{se^{-T_1}}). \]  

\begin{figure}[ht]
\begin{center}
\begin{minipage}{0.8\linewidth}
   \centering
   \includegraphics[angle=0,width=0.7\linewidth]{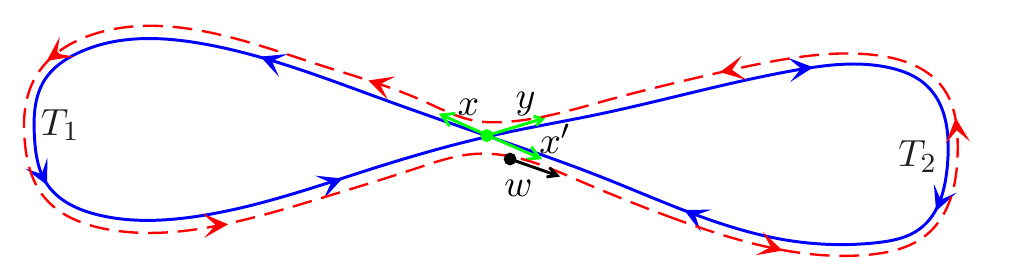}
\end{minipage}
\end{center}
\caption{Periodic partner orbit}\label{mainthm}
\end{figure}

\noindent
A short calculation using $s=\tan(\phi/2)$ reveals that 
\begin{eqnarray*} 
   \lefteqn{A_{T_2} B_{-s} D_\phi A_{-T_2} D_\phi B_{se^{-T_1}}}
   \\ & = & \Bigg(\begin{array}{cc} 1 & (1+e^{-T_1})\tan(\phi/2)
   \\[1ex] -(1+e^{-T_2})\sin(\phi/2)\cos(\phi/2) & 
   1-(1+e^{-T_2}+e^{-T_1}+e^{-T})\sin^2(\phi/2)\end{array}\Bigg)
   \\ & = & C_{\hat{u}} B_{\hat{s}},
\end{eqnarray*} 
where (see lemma \ref{decompo}(b))  
\[ \hat{s}=(1+e^{-T_1})\tan(\phi/2),\quad\hat{u}=-(1+e^{-T_2})\sin(\phi/2)\cos(\phi/2). \] 
Then 
\begin{eqnarray} |\hat{s}|&=&\frac{1+\sin^2(\phi/2)}{\cos(\phi/2)}\sin(\phi/2)<\frac32\sin(\phi/2)=\eps,
\\ |\hat{u}|&=& (1+\sin^2(\phi/2))\sin(\phi/2)<\frac32\sin(\phi/2)=\eps, 
\end{eqnarray}
by (\ref{tausu-bd}), (\ref{sinphi/2}), 
(\ref{cosphi/2}) and corollary \ref{periodandangle}. As a consequence of
(\ref{sinphi/2}), we have 
$\eps<\min\{\frac{1}{4},\frac{\sigma_0}{6}\}$.  Due to $\hat{y}=\Gamma g b_{s} a_{-T_2}$ therefore 
\[ \varphi^X_T(\hat{y})=\Gamma g b_{s} a_{-T_2} (c_{\hat{u}} b_{\hat{s}})
   \in {\cal P}_{\eps}(\hat{y}). \]
Furthermore, by the assumption $T\geq 1$, we can apply the Anosov closing lemma to have
$w\in {\cal P}_{2\eps}(\hat{y})$ and $T'\in\R$ so that 
\begin{equation}\label{jamkru} 
   \varphi^X_{T'}(w)=w\quad\mbox{and}\quad
   d_X(\varphi^X_t(\hat{y}), \varphi^X_t(w))< 4\eps
   \quad\mbox{for all}\quad t\in [0, T].
\end{equation} 
Furthermore, 
\begin{equation}\label{lyrekn} 
   \Big|\frac{T'-T}{2}-\ln(1+\hat{s}\hat{u})\Big|<5\,\eps^2 e^{-T}
   =\frac{45}{4}\,\sin^2(\phi/2) e^{-T}
\end{equation}
and 
\[ e^{T'/2}+e^{-T'/2}=e^{T/2}+e^{-T/2}+\hat{s}\hat{u}\,e^{T/2}; \] 
note that $|\hat{s}\hat{u}|<\eps^2<1$, i.e., $\ln(1+\hat{s}\hat{u})$ is well-defined. 
Since in fact  
\begin{equation}\label{hatsu7} 
   \hat{s}\hat{u}=-(1+e^{-T_1})(1+e^{-T_2})\sin^2(\phi/2)<0, 
\end{equation} 
we obtain  $T'<T$. 
Also (\ref{actdiff}) follows from (\ref{lyrekn}) and (\ref{hatsu7}). 
To prove that the orbits remain close, if $t\in [0, T_2]$, then by (\ref{jamkru}), 
(\ref{goverACL}), (\ref{fste3n}), and (\ref{versush}), 
\begin{eqnarray}\label{coe1} 
   d_X(\varphi^X_t(w), \varphi^X_t(x')) & \le & d_X(\varphi^X_t(w), \varphi^X_t(\hat{y}))
   +d_X(\varphi^X_t(\hat{y}), \varphi^X_t(x'))
   \nonumber
   \\ & < & 4\eps+d_X(\varphi^X_{t-T_2}(\hat{x}), \varphi^X_{t-T_2}(y'))<6\eps=9|\sin(\phi/2)|.  
\end{eqnarray} 
Similarly for $t\in [T_2, T]$ by (\ref{jamkru}), (\ref{fste2n}), and (\ref{shadcons1}), 
\begin{eqnarray*} 
   d_X(\varphi^X_t(w), \varphi^X_t(y)) 
   & \le & d_X(\varphi^X_t(w), \varphi^X_t(\hat{y}))+d_X(\varphi^X_t(\hat{y}), \varphi^X_t(y))
   \\ & \le & 4\eps+d_X(\varphi^X_{t-T_2}(\hat{x}), \varphi^X_{T_2-t}(x))< 4\eps+\eps e^{t-T_2}
   \le 5\eps<9|\sin(\phi/2)|. 
\end{eqnarray*} 
Hence if $t\in [0, T_1]$, then by (\ref{fste2n}), 
\begin{equation}\label{coe2} 
   d_X(\varphi^X_{t+T_2}(w), \varphi^X_t(x))=d_X(\varphi^X_{t+T_2}(w), \varphi^X_{t+T_2}(y))
   <9|\sin(\phi/2)|.
\end{equation} 
Defining $(q, \zeta)=\Xi^{-1}(w)\in {\cal X}=T^1(\Gamma\backslash\H^2)$, 
the orbit of this point is $T'$-periodic and it will have the desired properties 
by (\ref{coe1}) and (\ref{coe2}). {\hfill$\Box$}\bigskip


\begin{remark}\rm 
In what follows, we will drop the superscript 
$X$ from ${(\varphi_t^X)}_{t\in \R}$ to simplify notation.
\smallskip 

\noindent (a) It is a consequence of (\ref{formT'}) that the period $T'$ of the partner orbit is determined by
\[ T'=2\,\arccosh\left(\frac{e^{T/2}+e^{-T/2}- (1+e^{-T_1})(1+e^{-(T-T_1)})e^{T/2}\sin^2(\phi/2)}2\right). \]

\smallskip
\noindent
(b) Owing to (\ref{actdiff}) we have 
\begin{equation}\label{actioncosin} 
   \Big|\frac{T'-T}{2}-\ln(1-\sin^2(\phi/2))\Big|\leq 4(e^{-T_1}+e^{-T_2})\sin^2(\phi/2).
\end{equation} 
This explains the approximation of length difference $\Delta L=\Delta T\approx 4\ln\cos(\phi/2)$ 
as was obtained by Braun et al.~\cite{braun2002}. Furthermore, by Taylor expansion, 
\begin{eqnarray*}
   \ln(1-\sin^2(\phi/2))=-\sin^2(\phi/2)-\frac{\sin^4(\phi/2)}{2}-\frac{\sin^6(\phi/2)}{3}+o(\sin^6(\phi/2)).
\end{eqnarray*}
If all elements in $\Gamma\setminus\{e\}$ are hyperbolic, then by the self-crossing property, 
corollary \ref{periodandangle}(b): 
\[ e^{-T_1}<\sin^2(\phi/2)\quad\mbox{ and}\quad e^{-T_2}<\sin^2(\phi/2). \]
Therefore it follows from (\ref{actioncosin}) that
\[ \frac{T'-T}{2}=-\sin^2(\phi/2)+O(\sin^4(\phi/2)) \] 
is the asymptotics for $\phi$ small. 
\smallskip

\noindent
(c) Let all $\gamma\in\Gamma\setminus\{e\}$ be hyperbolic. Recall that $\tilde y=\Gamma h'a_{-\tau}=\varphi_{-\tau}(y')$. By the shadowing lemma, 
\[ d_X(\varphi_t(x),\varphi_t(\hat x))<\eps e^{-t}\quad\mbox{for all}\quad t\in [0,\infty[ \]
and
\[ d_X(\varphi_t(\tilde y),\varphi_t(\hat x))<\eps e^t\quad\mbox{for all}\quad
   t\in\,[-\infty, 0]. \] 
Furthermore, according to the proof of the Anosov closing lemma, 
\begin{eqnarray*}
   d_X(\varphi_t(w)),\varphi_t(\hat y))<2\eps e^{-t}+2\eps e^{t-T}\quad\mbox{for all}\quad t\in [0,T].
\end{eqnarray*}
Denoting $\tilde x=\Gamma g' a_{-\tau}=\varphi_{-\tau}(x')$, thus $\varphi_{T_2}(\tilde x)=\tilde y$
and by (\ref{shadcons2}) we have
\begin{eqnarray}\label{buncheq}
   \notag
   d_X(\varphi_t(w),\varphi_t(\tilde x))&\leq& d_X(\varphi_t(w),\varphi_t(\hat y))
   +d_X(\varphi_t(\hat y),\varphi_t(\tilde x))
   \\ \notag
   & \leq & 2\eps e^{-t}+2\eps e^{t-T}+ d_X(\varphi_{t-T_2}(\hat x),\varphi_{t-T_2}(\tilde y))
   \\ \notag
   & < & 2\eps e^{-t}+2\eps e^{t-T} + \eps e^{t-T_2}
   \\ & < & 2\eps (e^{-t}+e^{t-T_2}) \quad \mbox{for all}\quad t\in [0,T_2],
\end{eqnarray}
due to $e^{-T_1}<\sin^2(\phi/2)<\frac{1}{2}$. Considering the right-hand side 
of (\ref{buncheq}), the function $f(t)=2\eps (e^{-t}+e^{t-T_2})$ is decreasing f
or $t\in [0, \frac{T_2}{2}]$, increasing for $t\in [\frac{T_2}2,T_2]$, and minimal at $t=\frac{T_2}{2}$.
In particular,
\[ d_X(\varphi_{T_2/2}(w),\varphi_{T_2/2}(\tilde x))<4\eps e^{-T_2/2}. \]
For $t\in [T_2,T]$, 
\begin{eqnarray*}
   d_X(\varphi_t(w),\varphi_t(y))
   & \leq & d_X(\varphi_t(w),\varphi_t(\hat y))+d_X(\varphi_t(\hat y),y)
   \\ & \leq & 2\eps e^{-t}+2\eps e^{t-T}+ d_X(\varphi_{t-T_2}(\hat x),\varphi_{t-T_2}(x))
   \\ & < & 2\eps e^{-t}+2\eps e^{t-T} +\eps e^{T_2-t}.
\end{eqnarray*}
Similarly, for $t\in [0,T_1]$, 
\begin{eqnarray*}
   d_X(\varphi_{t+T_2}(w),\varphi_t(x))
   & = & d_X(\varphi_{t+T_2}(w),\varphi_{t+T_2}(y))
   <2\eps e^{-T_2-t}+2\eps e^{t-T_1}+\eps e^{-t}
   \\ & < & 2\eps (e^{-t}+e^{t-T_1})
\end{eqnarray*}
and the function $g(t)=e^{-t}+e^{t-T_1}$ attains the minimum at $t=\frac{T_1}2$:
\[ d_X(\varphi_{T_1/2+T_2}(w),\varphi_{T_1/2}(y))<3\eps e^{-T_1/2}. \]
This means that outside the encounter region, the partner orbits remain very close.
\smallskip

\noindent
(d) Recall $\gamma_1$ and $\gamma_2$ from (\ref{flukn}):
$\gamma_1=ga_{T_1}h^{-1}$ and $\gamma_2=ha_{T_2}g^{-1}$. 
Since $x=\Gamma g$ is a $T$-periodic orbit, we have 
\[ \gamma =ga_Tg^{-1} \]
for some $\gamma\in\Gamma$,
and by lemma \ref{exgeodper}, the orbit through $x$ (called  $c$) 
corresponds to the conjugacy class ${\{\gamma\}}_\Gamma$:
\[ \gamma_1\gamma_2=(ga_{T_1}h^{-1})(ha_{T_2}g^{-1})=ga_{T}g^{-1}=\gamma. \]
This means that the orbit $c$ corresponds to the conjugacy class ${\{\gamma\}}_\Gamma
={\{\gamma_1\gamma_2\}}_\Gamma$. On the other hand the partner orbit (called $c'$) corresponds to the conjugacy class
${\{\gamma'\}}_\Gamma={\{\gamma_1\gamma_2^{-1}\}}_\Gamma$. Indeed, according to the proof of Anosov closing lemma, 
the partner orbit $c'$ corresponds to the conjugacy class of
$\zeta^{-1}$ where $\zeta=\hat g c_{\hat u}b_{\hat s}a_{-T}{\hat g}^{-1}$ for
$\hat g=gb_sa_{-T_2}$. 
Using $h'=gd_\phi=gb_sc_ua_\tau$ and $h=g'd_\phi=g'a_{-\tau}c_ub_s$ we obtain 
\begin{eqnarray*}
   \zeta & = & \hat g c_{\hat u}b_{\hat s} a_{-T}{\hat g}^{-1}
   =gb_sa_{-T_2}c_{\hat u}b_{\hat s}a_{-T} a_{T_2}b_{-s}g^{-1}
   \\ & = & gb_sa_{-T_2}c_{(1+e^{-T_2})u}b_{(1+e^{-T_1})s}a_{-T_1}b_{-s}g^{-1}
   \\ & = & gb_sa_{-T_2}c_{(1+e^{-T_2})u}b_{(1+e^{-T_1})s}b_{-se^{-T_1}}a_{-T_1}g^{-1}
   \\ & = & h'a_{-\tau}c_{-u}a_{-T_2}c_{(1+e^{-T_2})u}b_{s}a_{-T_1}g^{-1}
   \\ & = & h'a_{-T_2}a_{-\tau}c_{-ue^{-T_2}} c_{(1+e^{-T_2})u}b_{s}a_{-T_1}g^{-1}
   \\ & = &\gamma_2 g'a_{-\tau}c_ub_sh^{-1}\gamma_1^{-1}=\gamma_2\gamma_1^{-1},
\end{eqnarray*}
noting
$h'a_{-T_2}=\gamma_2g'$ owing to $ha_{T_2}=\gamma_2 g$.
Therefore $\gamma'=\zeta^{-1}=\gamma_1\gamma_2^{-1}$ and we obtain the result by Braun et al.~\cite{braun2002}.
\smallskip

\noindent
(e) We are going to construct a periodic orbit  whose period is approximately $2T$. Denote 
\[ v=\varphi_{T_2/2}(w)=\Gamma g'a_\tau c_{ue^{-T_2}+\sigma}b_\eta a_{T_2/2}
    =\Gamma g'a_{\tau+T_2/2}c_{u e^{-T_2/2}+\sigma e^{T_2/2}}b_{\eta e^{-T_2/2}}
    =\Gamma g'a_{\tau+T_2/2}c_{\tilde u}b_{\tilde s}\]
for $\tilde u=u e^{-T_2/2}+\sigma e^{T_2/2}$ and $\tilde s=\eta e^{-T_2/2}$. We have 
\begin{eqnarray*}
   |\tilde u|&\leq& |u| e^{-T_2/2}+2|\hat u|e^{-T-T_2/2}<\frac{3}{2}\,|u|e^{-T_2/2}
   <\frac{3}{2}\,\eps e^{-T_2/2}=:\hat \eps,
   \\ |\tilde s| & \leq & \frac32\,|\hat s|e^{-T_2/2}<\hat{\eps}.
\end{eqnarray*} 
Write $v=\Gamma k$ for some $k\in\G$, then 
$\Gamma kb_{-\tilde s}=\Gamma g' a_{\tau+T_2/2}c_{\tilde u}=:z$. 
Next we apply the shadowing lemma to obtain 
\begin{eqnarray}\label{zv}
   d_X(\varphi_t(z),\varphi_t(v))<\hat \eps e^{-t}
   & & \mbox{for all}\quad t\in [0, \infty[
   \\ \label{zw} 
   d_X(\varphi_t(z),\varphi_t(\check x))<\hat \eps e^{t}
   & & \mbox{for all}\quad t\in\, ]-\infty, 0]
\end{eqnarray}
for $\check x=\Gamma g' a_{\tau+T_2/2}$. 
Denoting $\hat z=\varphi_{-T}(z)$, we can check that
\[ \varphi_{T+T'}(\hat z)=\Gamma (g' a_{\tau+T_2/2})c_{\hat u}b_{\hat s} \]
for 
\[ \check u=\tilde u(1-e^{-T})\quad\mbox{and}\quad \check s=-\tilde s(1-e^{-T'}). \]
This means that $\varphi_{T+T'}(\hat z)\in \P_{\hat \eps}(\hat z)$.
Now we utilize the Anosov closing lemma to find a $\hat T$-periodic point $z_*$ so that
\begin{equation}\label{zt} 
   d_X(\varphi_t(z_*),\varphi_t(\hat z))<4\hat\eps\quad\mbox{for all}\quad t\in [0, T+T']
\end{equation} 
and 
\begin{eqnarray}\label{hatT-}
   \Big|\frac{\hat T-(T+T')}2-\ln(1+\check u\check s)\Big|
   <5|\check u\check s|e^{-T-T'}.
\end{eqnarray} 
For $t\in [0,T]$, by (\ref{zv}) and (\ref{zt}):
\begin{eqnarray*}
   d_X(\varphi_t(z_*),\varphi_t(\check x))
   & \leq & d_X(\varphi_t(z_*),\varphi_t(\hat z))
   +d_X(\varphi_t(\hat z),\varphi_t(\check x))
   \\ & = & d_X(\varphi_t(z_*),\varphi_t(\hat z))
   +d_X(\varphi_{t-T}(z),\varphi_{t-T}(\check z))
   <5\hat \eps.
\end{eqnarray*}
For $t\in [0,T']$, by (\ref{zw}) and (\ref{zt}): 
\begin{eqnarray*}
   d_X(\varphi_{t+T}(z_*),\varphi_t(v))
   & \leq & d_X(\varphi_{t+T}(z_*),\varphi_{t+T}(\hat z))
   +d_X(\varphi_{t+T}(\hat z),\varphi_t(v))
   \\ & = & d_X(\varphi_{t+T}(z_*),\varphi_{t+T}(\hat z))
   +d_X(\varphi_{t}(z),\varphi_t(v))<5\hat \eps.  
\end{eqnarray*}
It follows from (\ref{hatT-}) that 
\[ \Big|\frac{\hat T-(T+T')}{2}\Big|<10\eps^2 e^{-T_2}. \]
Therefore
\begin{eqnarray*}
   \Big|\frac{\hat T -2T }{2} -\ln(1+\hat u\hat s)\Big|\leq \Big |\frac{\hat T-(T+T')}2 \Big|
   +\Big |\frac{T'-T}{2}-\ln(1+\hat u\hat s)\Big |<11\eps^2 e^{-T_2}
\end{eqnarray*}
by (\ref{lyrekn}), and we have found a periodic orbit whose the period $\hat T$ is close to $2T$; 
see Figure \ref{longorbit}.

\begin{figure}[ht]
\begin{center}
\begin{minipage}{0.8\linewidth}
   \centering
   \includegraphics[angle=0,width=0.7\linewidth]{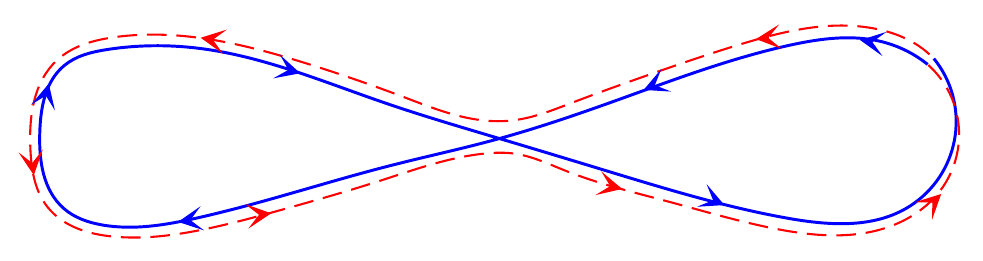}
\end{minipage}
\end{center}
\caption{Reconnect of a Sieber-Richter pair to form a longer orbit}\label{longorbit}
\end{figure}

\noindent
In the same way one can construct periodic orbits with periods approximately $nT$ for any $n\in\N$ 
which are close to the original orbit.
{\hfill$\diamondsuit$}
\end{remark}

From theorem \ref{existthm} one can also derive an $\eps$-$\delta$-version. 

\begin{theorem}[Existence of a partner orbit II] 
For every $\eps>0$ there is a $\delta>0$ with the following property.
If a periodic orbit of the geodesic flow on $T^1(\Gamma\backslash\H^2)$
with the period $T\geq 1$ crosses itself in configuration space at a time $T_1\in\,]0,T[$
and at an angle $\theta$ such that {$0<\phi<\delta$}  
for $\phi=\pi-\theta$, then there is another periodic orbit of the geodesic flow 
(called a partner orbit) which remains $\eps$-close to the original one. 
Furthermore, $T'<T$ for the period of the partner orbit, and 
\begin{eqnarray*}
   \Big|\frac{T'-T}{2}-\ln\Big(1-(1+e^{-T_1})(1+e^{-(T-T_1)})\sin^2(\phi/2)\Big)\Big|\leq \eps^2 e^{-T}.
\end{eqnarray*}
\end{theorem}
\noindent
{\bf Proof\,:} Fix $\eps>0$ and define 
\[ \delta=\min\Big\{\frac{2\eps}{9}, \frac{1}{3}\Big\}. \] 
If $|\phi|\le\delta$, then $9|\sin(\phi/2)|\le \frac{9|\phi|}2\le\eps$ 
and $12\sin^2(\phi/2)\le 3\phi^2\le\eps^2$. Therefore theorem \ref{existthm} applies. 
{\hfill$\Box$}\bigskip

\subsection{Uniqueness of the partner orbit}

In the compact case owing to the hyperbolicity two periodic orbits with similar periods 
cannot stay too close together without being identical. 

\begin{lemma}\label{per-coinc} Let $X=\Gamma\backslash {\rm PSL}(2, \R)$ be compact. 
Then there is $\eps_\ast>0$ with the following property. 
If $\eps\in \,]0, \eps_\ast[$ and if $x_1, x_2\in X$ 
are periodic points of ${(\varphi_t^X)}_{t\in\R}$ having the periods $T_1, T_2>0$ 
such that $|T_1-T_2|\le\sqrt 2\eps$ and 
\[ d_X(\varphi_t^X(x_1), \varphi_t^X(x_2))<\eps
   \quad\mbox{for all}\quad t\in [0, \min\{T_1, T_2\}], \]  
then $T_1=T_2$ and the orbits of $x_1$ and $x_2$ under ${(\varphi_t^X)}_{t\in\R}$ are identical. 
\end{lemma} 
\noindent 
{\bf Proof\,:} According to lemma \ref{injrad2} there is $\sigma_0>0$ 
such that $d_{\G}(u, \gamma u)\ge\sigma_0$ holds for all $u\in \G$ 
and $\gamma\in\Gamma\setminus\{e\}$.
In addition, lemma \ref{konvexa}(b) for $\eps=1$ 
implies that there is $\sigma_1>0$ with the following property. 
If $d_{{\rm PSL}(2, \R)}(u, e)<\sigma_1$, then there is 
$U=\scriptsize\Big(\begin{array}{cc} u_{11} & u_{12}
   \\ u_{21} & u_{22}\end{array}\Big)\in {\rm SL}(2, \R)$
such that $u=\pi(U)$ and $|u_{11}-1|+|u_{12}|+|u_{21}|+|u_{22}-1|<1$. 
Put $\eps_\ast=\min\{\frac{\sigma_0}{6}, \frac{\sigma_1}{2}\}$ 
and fix $\eps\in\, ]0, \eps_\ast[$. W.l.o.g.~we can assume that $T_1\le T_2$, 
and we write $c_j(t)=\varphi_t^X(x_j)$ for $j=1, 2$. 
Then $d_X(c_1(t), c_2(t))<\eps$ for $t\in [0, T_1]$ and $c_j$ 
is $T_j$-periodic. Let $g_1, g_2\in \G$ be such that 
$x_j=\Gamma g_j=\Pi_\Gamma(g_j)$ for $j=1, 2$.
Now defining $z_j(t)=c_j(tT_j)$, we see that $z_j$ is $1$-periodic. 
Furthermore, if $t\in [0, 1]$, then 
\begin{eqnarray*} 
   d_X(z_1(t), z_2(t)) & = & d_X(c_1(tT_1), c_2(tT_2)) 
    \le  d_X(c_1(tT_1), c_2(tT_1))+d_X(c_2(tT_1), c_2(tT_2))
   \\
   &\leq & \eps+d_{\G}(a_{tT_1}, a_{tT_2})
    \le  \eps+\frac{t}{\sqrt{2}}\,|T_1-T_2|\le 2\eps. 
\end{eqnarray*} 
Since both $z_1$ and $z_2$ are $1$-periodic, it follows that 
\begin{equation}\label{topfack} 
   d_X(z_1(t), z_2(t))<2\eps\quad\mbox{for}\quad t\in\R. 
\end{equation} 
Denoting $v_j(t)=g_j a_{tT_j}\in \G$ for $i=1, 2$, we obtain $\Pi_\Gamma(v_j(t))
=\Gamma g_j a_{tT_j}=\varphi_{tT_j}^X(x_j)=c_j(tT_j)=z_j(t)$. 
Hence by (\ref{topfack}) and the definition of $d_X$, 
for every $t\in\R$ there is $\gamma(t)\in\Gamma$ so that 
$d_{\G}(v_1(t), \gamma(t)v_2(t))<2\eps$ for $t\in\R$. 
It follows that $\gamma(t)=\gamma(0)$ for all $t\in\R$. 
Hence denoting $\gamma_0=\gamma(0)\in\Gamma$, we obtain 
\[ d_{\G}(a_{-tT_2} (\gamma_0 g_2)^{-1} g_1 a_{tT_1}, e)
   =d_{\G}(g_1 a_{tT_1}, \gamma_0 g_2 a_{tT_2})<2\eps<\sigma_1
   \quad\mbox{for\,\,all}\quad t\in\R. \] 
Write $(\gamma_0 g_2)^{-1}g_1=\pi(C)$ for
 \[ C=\Bigg(\begin{array}{cc} a & b
   \\ c & d\end{array}\Bigg)\in {\rm SL}(2, \R). \] 
Then $A_{-tT_2}CA_{tT_1}=\scriptsize \Big(\begin{array}{cc} e^{t(T_1-T_2)/2}a & e^{-t(T_1+T_2)}b 
\\ e^{t(T_1+T_2)/2}c & e^{-t(T_1-T_2)/2}d\end{array}\Big)$
and $\pi(\pm A_{-tT_2}CA_{tT_1})=a_{-tT_2} (\gamma_0 g_2)^{-1} g_1 a_{tT_1}$. 
Hence the definition of $\sigma_1$ leads to 
$e^{-t(T_1+T_2)}|b|+e^{t(T_1+T_2)/2}|c|<1$ for $t\in\R$.  
As $t\to\pm\infty$ we obtain $b=c=0$, and accordingly $ad=1$. 
In addition, for every $t\in\R$ we have either 
\begin{equation}\label{wildent1} 
   |e^{t(T_1-T_2)/2}a-1|+|e^{-t(T_1-T_2)/2}d-1|<1
\end{equation} 
or 
\begin{equation}\label{wildent2} 
   |-e^{t(T_1-T_2)/2}a-1|+|-e^{-t(T_1-T_2)/2}d-1|<1.
\end{equation} 
Suppose that $T_1\neq T_2$, and then $T_1<T_2$.
If (\ref{wildent1}) holds along a sequence $t=n_j\to\infty$, 
then $|e^{n_j(T_2-T_1)/2}d-1|<1$ leads to a contradiction. 
Similarly, if (\ref{wildent2}) holds along a sequence $t=n_j\to\infty$, 
then $|e^{n_j(T_2-T_1)/2}d+1|<1$ is impossible as $j\to\infty$. 
As a consequence, we must have $T_1=T_2$, and furthermore, 
due to $ad=1$, we can write either $a=e^\tau$, $d=e^{-\tau}$ 
or $a=-e^\tau$, $d=-e^{-\tau}$ for some $\tau\in\R$. 
Then $(\gamma_0 g_2)^{-1}g_1=\pi(C)=a_\tau$ 
shows that $g_1=\gamma_0 g_2 a_\tau$, and therefore 
\begin{eqnarray*} 
   \{\varphi_t^X(x_1): t\in\R\} & = & \{\Pi_\Gamma(g_1 a_t): t\in\R\}
   =\{\Pi_\Gamma(\gamma_0 g_2 a_\tau a_t): t\in\R\}
   =\{\Pi_\Gamma(g_2 a_s): s\in\R\}
   \\ & = & \{\varphi_s^X(x_2): s\in\R\}
\end{eqnarray*} 
for the orbits. 
{\hfill$\Box$}\bigskip

\begin{theorem}\label{uniq_thm}
In the setting of theorem \ref{existthm}, if $\Gamma\backslash\H^2$ is compact 
and the crossing angle $\phi$ satisfies
\[ |\phi|<\phi_0:=\min\Big\{\frac13,\frac{\eps_*}{9}\Big\}, \] 
then the partner orbit is unique; recall the number $\eps_*$ from lemma \ref{per-coinc}. 
\end{theorem}
\noindent
{\bf Proof\,:} If $\phi<\min\{\frac13,\frac{\eps_*}{9}\}$
then $\phi<\min\{\frac13,\frac{\sigma_0}6\}$. By theorem \ref{existthm} 
there exists a periodic orbit which is $9|\sin(\phi/2)|$-close to the original one 
and its period $T'$ satisfies $|T'-T|<4\sin^2(\phi/2)$. 
Assume that there is another partner orbit which has the same property, i.e., 
it is also $9|\sin(\phi/2)|$-close to the original orbit and its period called $T''$ satisfies
$|T''-T|<4\sin^2(\phi/2)$. Then these two partner orbits are $18|\sin(\phi/2)|$-close to each other and their periods satisfy
\[ |T''-T'|\leq |T''-T|+|T'-T|<8\sin^2(\phi/2)<\eps_*. \]
Due to $\phi<\min\{\frac13,\frac{\eps_*}{9}\}$ we obtain 
$18|\sin(\phi/2)|<\eps_*$. 
Therefore these two partner orbits must be identical, 
as a consequence of lemma \ref{per-coinc}. 
{\hfill$\Box$}\bigskip

\subsection{Encounter duration}

Consider the setting of theorem \ref{existthm} and suppose that $\Gamma\backslash\H^2$ 
is compact and the crossing angle satisfies $|\phi|<\phi_0=\min\{1/6,\eps_*/9\}$ as the preceding theorem.   
Then $\eps=\frac{3}{2}\sin(\phi/2) <\frac{\eps_*}{12}=:\varrho$. 
Recall from the proof of theorem \ref{existthm} that $x=\Gamma h'a_{-\tau}c_{-u}b_{-s}\in \P_\varrho(\tilde y)$, 
where $u=-\sin(\phi/2)\cos(\phi/2)$ and $s=\tan(\phi/2)$. Using $b_{-s}a_t=a_tb_{-se^{-t}}$ 
and $c_{-u}a_t=a_tc_{-u e^t}$, we obtain 
\[ \varphi_t(x)=\Gamma h' a_{-\tau}a_t c_{-ue^{t}}b_{se^{-t}}\in \P_\varrho(\varphi_t(\tilde y)) \mbox{
 if and only if } |u|e^{t}<\varrho \mbox{ and } |s|e^{-t}<\varrho,\]
  or equivalently, 
\[-\ln\Big(\frac{\varrho}{|s|}\Big)<t<\ln\Big(\frac{\varrho}{|u|}\Big).\]
Then for $t\in\,]-\ln(\frac{\varrho}{|s|}),\ln(\frac{\varrho}{|u|})[$, we have
$\varphi_t(x)\in \P_\varrho(\varphi_t(\tilde y))$ and hence
$d_X(\varphi_t(x),\varphi_t(\tilde y))<2\varrho$. The encounter duration is thus given by 
\[ t_{{\rm enc}}=t_s+t_u=\ln\Big(\frac{\varrho^2}{|us|}\Big)
   =\ln\Big(\frac{\varrho^2}{\sin^2(\phi/2)}\Big), \]
where
$t_s=\ln(\frac{\varrho}{\tan(\phi/2)})$ and $t_u=\ln(\frac{\varrho}{\sin(\phi/2)\cos(\phi/2)})$ 
are the times that the original orbit can go backward, respectively forward, 
from $x$ before leaving the encounter region (see Figure \ref{mainthm}).
We see that the smaller the crossing angle $\phi$ is, the longer will be the encounter duration.
Noting that $\sin^2(\frac{\phi}2)<\frac{4}9\varrho^2$, the encounter duration has a uniform lower bound: 
\[t_{\rm enc}>\ln 9-\ln 4.\]

\subsection{Pseudo-orbits} 

\begin{definition}
Let $\eps>0$ be small. 

\noindent
\smallskip 
(a) 
We say that two given periodic orbits of the flow ${(\varphi^X_t)}_{t\in\R}$ 
have the {\em $\eps$-property} if there is a point on one orbit belonging 
to the Poincar\'e section of radius $\eps$ at some point on the other orbit.

\smallskip 
\noindent
(b) We say that $n$ given periodic orbits $c_1, \ldots, c_n$ of the flow
${(\varphi^X_t)}_{t\in\R}$ are an {\em $\eps$-chain} 
if each couple $c_j$ and $c_{j+1}$, $j=1, \ldots, n-1$, has the $\eps$-property. 

\smallskip 
\noindent
(c) Two given periodic orbits $c$ and $c'$ 
of the flow ${(\varphi^X_t)}_{t\in\R}$ are called {\em connected} 
if there is a chain $c_1, \ldots,c_n$ such that $c, c_1, \ldots, c_n, c'$ create an $\eps$-chain. 
In this case we say that $c$ and $c'$ are connected by the chain $c_1, \ldots, c_n$. 
\end{definition}
\begin{figure}[ht]
\begin{center}
\begin{minipage}{0.55\linewidth}
   \centering
   \includegraphics[angle=0,width=0.7\linewidth]{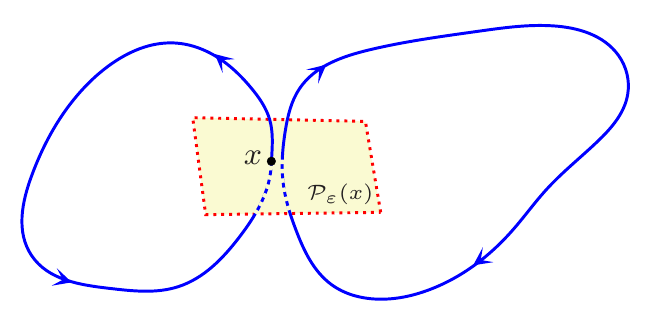}
\end{minipage}
\end{center}
\caption{Two periodic orbits having the $\eps$-property create a pseudo-orbit}\label{epsilon}
\end{figure}

\begin{definition}[Pseudo-orbit]
(a) A {\em pseudo-orbit} of the flow ${(\varphi_t^X)}_{t\in\R}$ is a finite set of periodic orbits 
in which any two elements are connected by an $\eps$-chain. We say that this orbit decomposes 
into periodic orbits. The period of a pseudo-orbit is the sum of the periods of its elements.  
The number of  elements which form a pseudo-orbit is called the rank of this pseudo-orbit.  

\smallskip 
\noindent
(b) A pseudo-orbit of the geodesic flow ${(\varphi_t^\X)}_{t\in\R}$ is a collection of periodic orbits which are
the image of a pseudo-orbit of the flow ${(\varphi_t^X)}_{t\in\R}$
under the isometry $\Xi^{-1}: X=\Gamma\backslash\PSL(2,\R) 
\rightarrow \X=T^1(\Gamma\backslash \H^2)$ from theorem (\ref{Xidef}). 
\end{definition}

\begin{theorem}
If a periodic orbit of the geodesic flow ${(\varphi^\X_t)}_{t\in\R}$ on $\X=T^1(\Gamma\backslash\H^2)$
with the period $T\ge 2$ crosses itself in configuration space at a time $T_1\in\,[1,T[$
and at an angle $\theta$ such that $\theta<\frac14$ and $T-T_1\ge 1$,
then there is a pseudo-orbit (called pseudo-partner) of the geodesic flow
which remains $8|\sin(\theta/2)|$-close to the original one.
Furthermore, $T'<T$ for the period of the pseudo-partner orbit, and 
\begin{eqnarray}\label{T'pseudo}
   \Big|\frac{T'-T}{2}-\ln(\cos^2(\theta/2))\Big|\leq 10\sin^2({\theta}/2)(e^{-T_1}+e^{-(T-T_1)}).
\end{eqnarray} 
\end{theorem}
\noindent
{\bf Proof\,:} Denote $T_2=T-T_1$. 
Let the orbit of $(p,\xi)\in\X=T^1(\Gamma\backslash\H^2)$ be $T$-periodic ($T$ being its prime period)
and such that it has a self-crossing of angle $\theta$ in configuration space at the time $T_1\in ]1,T[$.
According to the proof of theorem \ref{existthm}, 
if $x=\Xi(p,\xi)$ and $y=\varphi_{T_2}^X(x)$ then we may write $x=\Pi_\Gamma(g)$ and
$y=\Pi_\Gamma(h)$ for some $g,h\in\PSL(2,\R)$ so that
either $g=hd_\theta$ or $h=gd_\theta$. Henceforth we are going to assume 
that $h=gd_\theta$ since the case where $g=hd_\theta$ can be treated analogously. 
Using lemma \ref{decompo}(c) we can write
\begin{equation}\label{dthetap}
   d_\theta=c_{u_1}b_{s_1}a_{\tau_1},
\end{equation}
where
\[ u_1=-\tan(\theta/2),\quad s_1=\sin(\theta/2)\cos( \theta/2),\quad\tau_1=2\ln(\cos(\theta/2)). \]
\begin{figure}[ht]
\begin{center}
\begin{minipage}{0.55\linewidth}
   \centering
   \includegraphics[angle=0,width=0.7\linewidth]{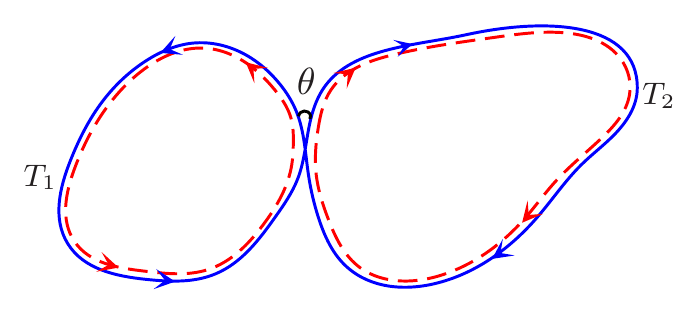}
\end{minipage}
\end{center}
\caption{A pseudo partner orbit}\label{pso}
\end{figure}
\noindent
Then 
\[ |u_1|\leq 2|\sin(\theta/2)|=:\eps,\quad |s_1|\leq |\sin(\theta/2)|<\eps, \]
and
\[ |\tau_1|=|\ln(1-\sin^2(\theta/2))|\leq 2\sin^2(\theta/2)|<\frac12\,\eps^2, \]
owing to $|\ln(1+z)|\leq 2|z|$ for $|z|<\frac{1}{2}$. Then
$\varphi_{T_1-\tau_1}(x)=\varphi_{-\tau_1}(y)=\Gamma gc_{u_1}b_{s_1}\in\P_\eps(x)$.
Hence, by the Anosov closing lemma, there are $x'=\Gamma gc_{\sigma_1} b_{\eta_1}\in\P_{2\eps}(x)$ 
and $T_1'\in\R$ so that $\varphi_{T_1'}(x')=x'$, 
\begin{equation}\label{pesx'}
   d_X(\varphi_t(x'),\varphi_t(x))\leq 4\eps
   =8\sin(\theta/2)\quad\mbox{for all}\quad t\in[0,T_1-\tau_1],
\end{equation}
and 
\begin{equation}\label{pseudo0}
\left|\frac{T_1'-(T_1-\tau_1)}{2}-\ln(\cos^2(\phi/2))\right|\leq 5|u_1s_1| e^{-T_1+\tau_1}<5\sin^2(\theta/2) e^{-T_1}. 
\end{equation}
Similarly, $x=\Gamma g=\Gamma hd_{-\theta}=\Gamma h c_{u_2}b_{s_2}a_{\tau_2}$ for
\[ u_2=\tan(\theta/2),\quad s_2=-\sin(\theta/2)\cos(\theta/2),\quad
   \tau_2=2\ln(\cos(\theta/2)). \]
It follows that $\varphi_{T_2-\tau_2}(y)=\varphi_{-\tau_2}(x)=\Gamma h c_{u_2}b_{s_2}\in\P_\eps(y)$.
Applying the Anosov closing lemma again, there are $y'=\Gamma hc_{\sigma_2}b_{\eta_2}\in\P_{2\eps}(y)$ 
and $T_2'\in\R$ such that $\varphi_{T_2'}(y')=y'$, 
\begin{equation}\label{pesy'}
   d_X(\varphi_t(y'),\varphi_t(y))\leq 4\eps=8\sin(\theta/2)\quad\mbox{for all}\quad t\in[0,T_2],
\end{equation}
and
\begin{equation}\label{pseudo2}
   \left|\frac{T_2'-(T_2-\tau_2)}{2}-\ln(\cos^2(\theta/2))\right|
   <5|u_2s_2|e^{-T_2+\tau_2}<5\sin^2(\theta/2) e^{-T_2}. 
\end{equation}
It follows from (\ref{pseudo0}) and (\ref{pseudo2}) that
\[ \left|\frac{(T'_1+T'_2)-T}{2}-\ln(\cos^2(\theta/2))\right| <10\sin^2(\theta/2)(e^{-T_1}+e^{-T_2}), \]
so that we obtain (\ref{T'pseudo}) for  $T'=T_1'+T_2'$. Since $u_1 s_1<0$ and $u_2s_2<0$, 
we have $T'_1<T_1$ as well as $T_2'<T_2$ and so $T'<T_1+T_2=T$. 
Defining $(q,\zeta)=\Xi^{-1}(x')\in {\cal X}$, $(l,\eta)=\Xi^{-1}(y')\in {\cal X}$, the orbit of
$(q,\zeta)$ is $T_1'$-periodic and the orbit of $(l,\eta)$ is
$T_2'$-periodic, and they will have the desired properties by (\ref{pesx'}) and (\ref{pesy'}). 
It remains to prove that the orbits of $x'$ and $y'$ create a pseudo-orbit.
Due to $h=gd_\theta$, (\ref{dthetap}), and using lemma \ref{decompo}(b), we obtain
\begin{eqnarray*}
   y' & = & \Gamma h c_{\sigma_2}b_{\eta_2}=\Gamma g d_\theta c_{\sigma_2}b_{\eta_2}
   =\Gamma g c_{u_1}b_{s_1}a_{\tau_1} c_{\sigma_2}b_{\eta_2}
   \\ & = & \Gamma (gc_{\sigma_1} b_{\eta_1})b_{-\eta_1}c_{-\sigma_1+u_1} 
   b_{s_1}c_{\sigma_2 e^{-\tau_1}}b_{\eta_2 e^{\tau_1}}a_{\tau_1}
   \\ & = & \Gamma (g c_{\sigma_1}b_{\eta_1})(c_ub_sa_\tau)
\end{eqnarray*}
for
\begin{eqnarray*}
   u & = & u_1-\eta_1+\sigma_2 e^{-\tau_1} +\frac{1}{1+\rho}((u_1-\sigma_1)\sigma_2 e^{-\tau_1}s_1
   -(u_1-\eta_1+\sigma_2 e^{-\tau_1})\rho),
   \\ s & = & s_1-\eta_1+\eta_2 e^{\tau_1} +\rho((2+\rho)\eta_2 e^{\tau_1}+s_1-\eta_1)-s_1\eta_1(u_1-\sigma_1)(1+\rho),
   \\ \tau & = & 2\ln(1+\rho)+\tau_1,
\end{eqnarray*}
where 
\[\rho= \sigma_2 e^{-\tau_1}(s_1-\eta_1)-s_1\eta_1(1+s_1\sigma_2e^{-\tau_1}).\]
A short calculation shows that $|u|<3\eps$ as well as $|s|<\eps$; 
hence $\varphi_{-\tau}(y')=\Gamma (g b_{\eta_1}c_{\sigma_1})c_{u}b_s\in\P_{3\eps}(x')$ 
and the orbits of $x'$ and $y'$ create a pseudo-orbit.
{\hfill$\Box$}\bigskip
 

\end{document}